\documentstyle{mn}\twocolumn

\newif\ifAMStwofonts

\def\mincir{\raise -2.truept\hbox{\rlap{\hbox{$\sim$}}\raise5.truept \hbox{$<$}\ }}
\def\mincireq{\hbox{\raise0.5ex\hbox{$<\lower1.06ex\hbox{$\kern-1.07em{\sim}$}$}}}
\def\magcir{\raise-2.truept\hbox{\rlap{\hbox{$\sim$}}\raise5.truept \hbox{$>$}\ }}

\title{Non-thermal emission in radio galaxy lobes: \,\\ 
II. Centaurus\,A, Centaurus\,B, and NGC\,6251}

\author[Persic \& Rephaeli]
       {Massimo Persic$^{1,2,3}$, 
        Yoel Rephaeli$^{4,5}$\\
        $^1$INAF-Trieste Astronomical Observatory, via G.B.\,Tiepolo 11, I-34100 Trieste, Italy \\
        $^2$INFN-Trieste, via A.\,Valerio 2, I-34127 Trieste, Italy \\
        $^3$Physics \& Astronomy Dept., Bologna University, via P.\,Gobetti 93/2, I-40129 Bologna, Italy \\
        $^4$School of Physics \& Astronomy, Tel Aviv University, Tel Aviv 69978, Israel \\
        $^5$Center for Astrophysics and Space Sciences, University of California at 
		San Diego, La Jolla, CA 92093, USA} 
\date{Accepted ... ... ... ... ;
      Received ... ... ... ... ;
      in original form ... ... ... ...}

\pubyear{2019}

\begin{document}

\maketitle

\label{firstpage}

\begin{abstract}

Radio and $\gamma$-ray measurements of large lobes of several radio galaxies provide adequate basis for 
determining whether emission in these widely separated spectral regions is largely by energetic electrons. 
This is very much of interest as there is of yet no unequivocal evidence for a significant energetic 
proton component to account for $\gamma$-ray emission by neutral pion decay. A quantitative assessment 
of the proton spectral distribution necessitates full accounting of the local and background radiation 
fields in the lobes; indeed, doing so in our recent analysis of the spectral energy distribution of the 
Fornax\,A lobes considerably weakened previous conclusions on the hadronic origin of the emission measured 
by the {\it Fermi} satellite. We present the results of similar analyses of the measured radio, X-ray and 
$\gamma$-ray emission from the lobes of Centaurus\,A, Centaurus\,B, and NGC\,6251. The results indicate 
that the measured $\gamma$-ray emission from these lobes can be accounted for by Compton scattering of the 
radio-emitting electrons off the superposed radiation fields in the lobes; consequently, we set upper bounds 
on the energetic proton contents of the lobes.
\end{abstract}

\begin{keywords}
galaxies: cosmic rays -- galaxies: active -- galaxies: individual: Centaurus\,A -- galaxies: individual: Centaurus\,B -- 
galaxies: individual: NGC\,6251 -- gamma rays: galaxies -- radiation mechanisms: non-thermal
\end{keywords}

\maketitle
\markboth{Persic \& Rephaeli: NT emission in the giant lobes of Cen\,A, Cen\,B, NGC\,6251}{}

\section{Introduction}

Measurements of non-thermal (NT) emission from the extended lobes of several nearby radio galaxies provide 
a basis for detailed modeling of the spectral distributions of energetic particles in these environments. 
Sampling the spectral energy distributions (SED), even with only limited spatial information, yields 
valuable insight on the emitting electrons and possibly also on energetic protons whose {\it p--p} interactions 
in the ambient lobe plasma and ensuing $\pi^0$-decay could yield detectable $\magcir 0.1$ GeV emission. In 
addition to the intrinsic interest in physical conditions in radio lobes, modeling energetic particles and 
their emission processes can yield important insight also for the origin of these particles in galaxy clusters.

Currently available spectral radio, X-ray, and $\gamma$-ray measurements of the lobes of the nearby galaxy 
Fornax\,A provide an adequate basis for determining the emission processes, the spectral energy distribution 
of the emitting particles, and the mean value the magnetic field, as has been done by McKinley et al. (2015), 
Ackermann et al. (2016), and more recently by us (Persic \& Rephaeli 2019; hereafter PR19). Here we carry out 
similar data analyses of three other radio galaxies whose multi-spectral lobe emission was detected (also) by 
{\it Fermi}-LAT: Centaurus\,A (Cen\,A), Centaurus\,B (Cen\,B), and NGC\,6251. Our improved spectral modeling 
of the measured SEDs of these sources is based on an updated EBL model (Franceschini \& Rodighiero 2017, 2018), 
and on a more complete accounting for the local (ambient) galactic radiation fields than done in previous works. 

In light of the fact that the results presented here are based on essentially identical treatment to that in 
PR19 (other than the spatially sectionalized modeling of Cen\,A), our discussion will be brief and limited only 
to the most relevant observational data and to the results of our new spectral modeling. In Sections 2 and 3 we 
summarize, respectively, the observational data and estimates of the radiation field densities in the lobes of 
the three galaxies. Results of the fitted SED models are detailed in Section 4 and discussed in Section 5. Our 
main conclusions are summarized in Section 6.

\section{Lobe SED measurements}

Here we summarize the measurements and main properties of the broad-band emission from the lobes of the radio 
galaxies studied in this paper.

\subsection{Cen\,A}

At a luminosity distance $D_L = 3.8$\,Mpc (Harris et al. 2010), Cen\,A has two giant elongated radio lobes, 
each with projected linear size $280$\,kpc by $140$\,kpc, centered nearly symmetrically to the north and south 
of the elliptical galaxy NGC\,5128. These were the first lobes detected by {\it Fermi}-LAT (Abdo et al. 2010). 
The improved spectral (0.06-30\,GeV) and spatial resolution attained in more recent radio, {\it Planck}, and 
{\it Fermi}-LAT measurements (Sun et al. 2016, hereafter SYMA16) allow a spatially resolved spectral analysis. 
On sufficiently large scale the radio and $\gamma$-ray emissions appear reasonably uniform across 
the lobes, but with a somewhat different N and S lobe morphology. X-ray observations and spectral 
analyses of Cen\,A were carried out by several groups (Abdalla et al. 2018 and references therein); 
however, due to the complex morphology of the source (e.g., Schreier et al. 1979; Hardcastle et al. 
2009 and references therein), there is no unambiguous detection of NT emission at $\sim$1\,keV (e.g., 
Kraft et al. 2002). Detailed lobe SED modeling (Abdo et al. 2010; Yang et al. 2012; SYMA16) 
suggested that the $\gamma$-ray emission from the lobes is Compton-upscattered radiation from the 
Cosmic Microwave Background (CMB) with additional contribution at higher energies from the 
Extragalactic Background Light (EBL).

\begin{figure}
\vspace{7.5cm}
\includegraphics{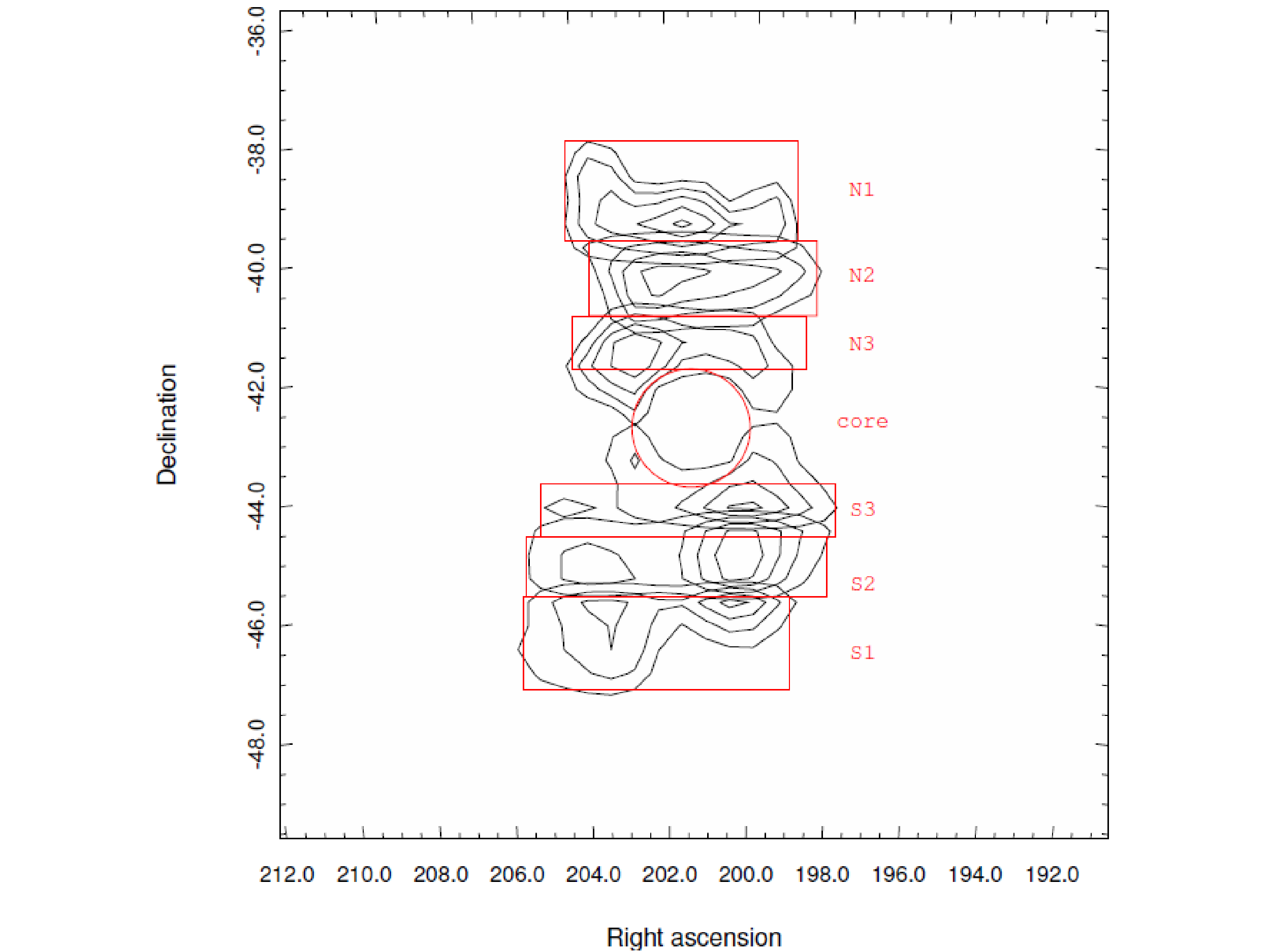}
\caption{
The lobes of Cen\,A (reproduced from Fig.\,3 of SYMA16, with permission). The contours correspond to the {\it 
Fermi}/LAT $\gamma$-ray ($>1$ GeV) image. The rectangles (red in the online version) indicate the regions of the 
radio and Plank aperture 
photometry, and the contours inside the corresponding regions show the template SYMA16 used for the extraction 
of the corresponding LAT spectrum. N1, N2, N3 are the outer, middle, inner regions of the northern lobe; and S1, 
S2, S3 are the outer, middle, inner regions of the southern lobe. The circle denotes the core region.
}
\label{fig:CenA}
\end{figure}

We use the most updated data in the radio and $\gamma$-ray bands, as reported by SYMA16: they are specified 
in Table\,1 where the listed flux densities are the emissions from the sub-regions in the north (N) and south (S) 
lobes identified by SYMA16 (see Fig.\,\ref{fig:CenA}) as N1, S1 (outer), N2, S2 
(middle), and N3, S3 (inner). Radio data comprise 118\,MHz Murchison Widefield Array (McKinley et al. 2013), 408\,MHz 
Haslem (Haslam et al. 1982), and 1.4\,GHz Parkes fluxes (O'Sullivan et al. 2013), whereas microwave data comprise 40, 
44, 70, 100, 143\,GHz {\it Planck} fluxes. These flux densities were measured using aperture photometry over the same 
sub-regions as specified in the analysis of the extended $\gamma$-ray image (SYMA16). As noted by SYMA16, for 
N2, S2, N3, and S3 the Planck data above 70 GHz may be affected by poor understanding of the high frequency background 
in this band hence may not be reliable. Based on 7\,yr of {\it Fermi}/LAT data and updated {\it Fermi}-LAT collaboration 
software tools, SYMA16 extended the detected $\gamma$-ray emission down to $60$ MeV and up to $30$ GeV; using an 
extended-lobe model for the lobe emission consistent with the radio lobes' morphology and point-like emission from the 
radio core, they confirmed Yang et al.'s (2012) previous finding that the North lobe's $\gamma$-ray emission extends 
beyond the radio image. 

\begin{table*}
\caption[] {Cen\,A: emission from the lobe regions.}
\begin{tabular}{ c  c c  r  r  r  r  r  r}
\hline
\hline
\noalign{\smallskip}
Frequency          & & &  N1~~~~~~  &   N2  ~~~~~~       &   N3 ~~~~~~       &    S1 ~~~~~~        &    S2  ~~~~~~      &     S3 ~~~~~~      \\
\noalign{\smallskip}
\noalign{\smallskip}
\cline{4-9}
\noalign{\smallskip}
  Log($\nu$/Hz)    & & &   \multicolumn{6}{|c|}{Radio Flux Density [Jy]$^{\dagger}$}                                                                   \\
\noalign{\smallskip}

$~\,8.072$ & & & $362.65 \pm 48.26$ & $517.63 \pm 61.86$ & $620.58 \pm 70.84$ & $479.99 \pm 60.31$ & $849.74 \pm 95.37$ & $764.29 \pm 86.30$ \\
$~\,8.611$ & & & $165.07 \pm 18.78$ & $265.98 \pm 28.95$ & $291.83 \pm 31.21$ & $368.12 \pm 39.64$ & $495.86 \pm 52.01$ & $382.24 \pm 40.48$ \\
$~\,9.146$ & & &  $72.25 \pm  1.55$ &  $97.56 \pm  2.04$ & $107.22 \pm  2.22$ &  $81.13 \pm  1.73$ & $163.92 \pm  3.37$ & $162.55 \pm  3.34$ \\
$10.477$   & & &   $9.63 \pm  0.01$ &  $11.19 \pm  0.01$ &  $12.41 \pm  0.01$ &   $9.29 \pm  0.01$ &  $19.26 \pm  0.01$ &  $24.36 \pm  0.01$ \\
$10.643$   & & &   $5.43 \pm  0.03$ &   $8.03 \pm  0.03$ &   $8.67 \pm  0.02$ &   -- ~~~~~~        &  $10.82 \pm  0.03$ &  $16.63 \pm  0.02$ \\
$10.845$   & & &   $2.37 \pm  0.07$ &  $11.70 \pm  0.06$ &   $7.92 \pm  0.05$ &   -- ~~~~~~        &  $15.67 \pm  0.06$ &  $12.58 \pm  0.05$ \\
$11.000$   & & &     -- ~~~~~~      &   $4.45 \pm  0.05$ &    -- ~~~~~~       &   -- ~~~~~~        &   $9.11 \pm  0.04$ &  -- ~~~~~~         \\
$11.155$   & & &     -- ~~~~~~      &    -- ~~~~~~       &    -- ~~~~~~       &   -- ~~~~~~        &   $3.92 \pm  0.04$ &  -- ~~~~~~         \\
\noalign{\smallskip}
\cline{4-9}
\noalign{\smallskip}
                   & & &    \multicolumn{6}{|c|}{$\gamma$-ray Flux Density [pJy]$^{\ddagger}$}                                                              \\
\noalign{\smallskip}
$22.50 \pm 0.125$ & & &  -- ~~~~~~        &  $4.70 \pm 0.50$  &       -- ~~~~~~   &       -- ~~~~~~   &        -- ~~~~~~   &       -- ~~~~~~  \\
$22.75 \pm 0.125$ & & &  $3.22 \pm 0.46$  &  $2.17 \pm 0.44$  &  $6.34 \pm 0.72$  &       -- ~~~~~~   &  $12.02 \pm 1.33$  &  $6.31 \pm 0.76$ \\
$23.00 \pm 0.125$ & & &  $1.10 \pm 0.37$  &  $1.33 \pm 0.35$  &  $2.40 \pm 0.44$  &  $1.20 \pm 0.44$  &   $5.25 \pm 0.67$  &  $2.09 \pm 0.45$ \\
$23.25 \pm 0.125$ & & &  $0.60 \pm 0.21$  &  $0.69 \pm 0.22$  &  $0.95 \pm 0.21$  &  $0.52 \pm 0.31$  &   $2.00 \pm 0.30$  &  $0.79 \pm 0.22$ \\
$23.50 \pm 0.125$ & & &  $0.48 \pm 0.12$  &  $0.21 \pm 0.09$  &  $0.34 \pm 0.11$  &  $0.39 \pm 0.12$  &   $1.00 \pm 0.14$  &  $0.29 \pm 0.10$ \\
$23.75 \pm 0.125$ & & &  $0.25 \pm 0.07$  &  $0.14 \pm 0.05$  &  $0.13 \pm 0.06$  &  $0.25 \pm 0.07$  &   $0.39 \pm 0.07$  &  $0.15 \pm 0.06$ \\
$24.00 \pm 0.125$ & & &  $0.07 \pm 0.04$  &  $0.12 \pm 0.04$  &  $0.12 \pm 0.04$  &  $0.13 \pm 0.04$  &   $0.15 \pm 0.04$  &  $0.07 \pm 0.04$ \\
$24.25 \pm 0.125$ & & &  $0.04 \pm 0.01$  &       -- ~~~~~~   &  $0.06 \pm 0.02$  &  $0.08 \pm 0.01$  &   $0.06 \pm 0.02$  &       -- ~~~~~~  \\
$24.50 \pm 0.125$ & & &  $0.03 \pm 0.02$  &       -- ~~~~~~   &  $0.03 \pm 0.02$  &       -- ~~~~~~   &   $0.04 \pm 0.02$  &       -- ~~~~~~  \\

\noalign{\smallskip}
\hline
\hline
\end{tabular}
\begin{flushleft}
\noindent
$^{\dagger}$From Table 3 of SYMA16. $^{\ddagger}$From Fig.\,4 of SYMA16.
\end{flushleft}
\end{table*}

\subsection{Cen\,B} 

\begin{figure}
\vspace{5.5cm}
\includegraphics{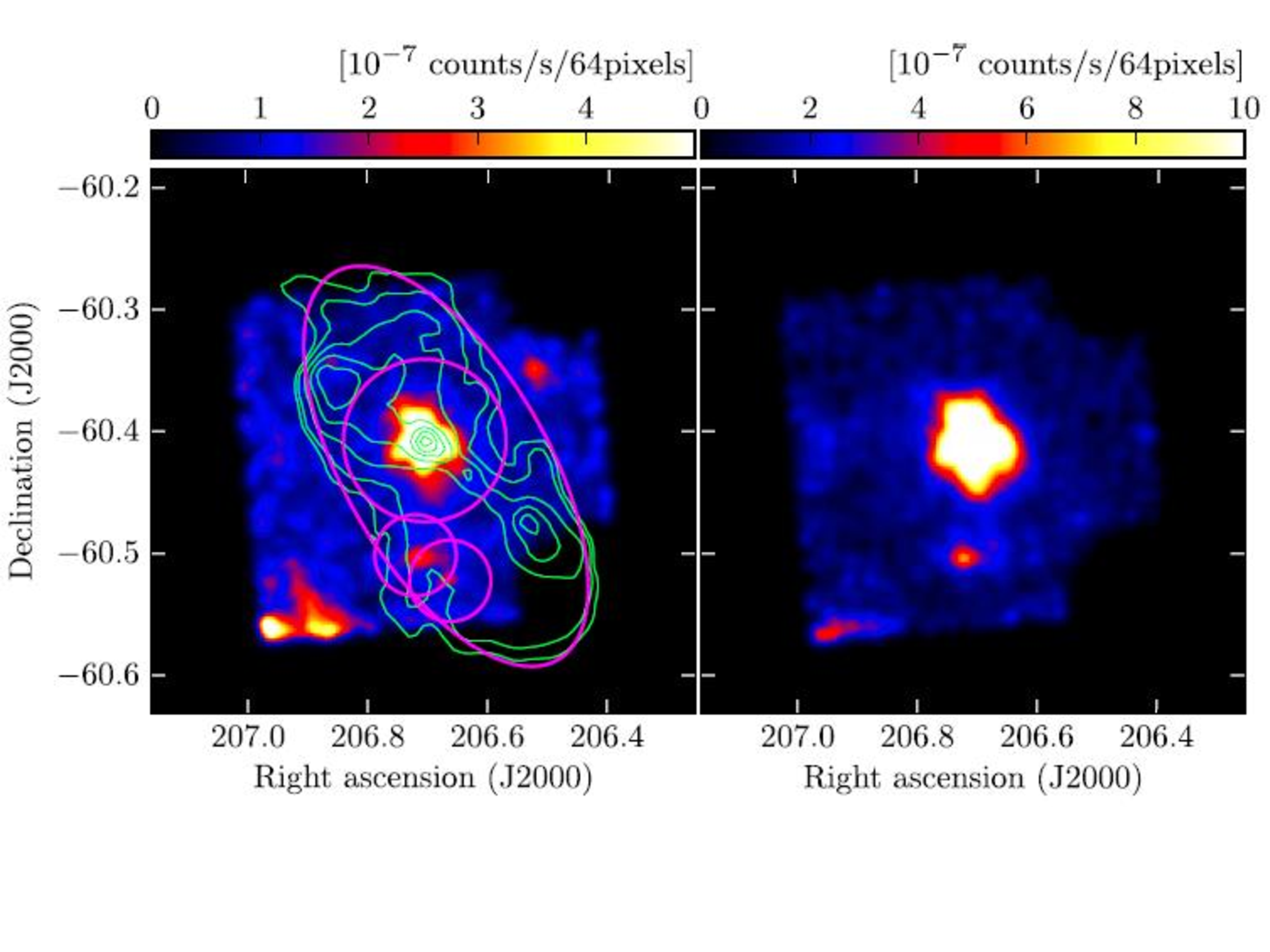}
\caption{
{\it Suzaku} image ({\it left}: 0.5-2 keV; {\it right}: 2-10 keV) of Cen\,B (reproduced from Fig.\,8 of K13, 
with permission). The central bright source is the nucleus of the system. The contours (green in the online 
version) denote the 843 MHz (McAdam 1991). The ellipse (magenta in the online version) denote the source 
extraction region for the lobes. The three circular regions (magenta in the online version) containing X-ray 
point-like sources were excluded from the K13 analysis. 
}
\label{fig:CenB}
\end{figure}

At $D_L = 56$\,Mpc, Cen\,B is the fifth-brightest radio galaxy in the sky. Its projected 
position close to the Galactic plane results in strong foreground absorption in the optical 
and X-ray bands (Schroeder et al. 2007; Tashiro et al. 1998). {\it Fermi}-LAT detecting emission 
from the lobes during $\sim 3.5$ yr observations (Katsuta et al. 2013, hereafter K13; see  
Fig.\,\ref{fig:CenB}). The level of diffuse X-ray emission from the lobes, 
originally claimed based on {\it ASCA} data (Tashiro et al. 1998), is marginally consistent with {\it 
Suzaku} upper limits (K13). If the flux is close to the {\it Suzaku} limit (which would be consistent 
with the {\it ASCA} estimate), then $\gamma$-ray emission could be interpreted as Compton scattering 
of the radio-emitting off the CMB; otherwise, a jet origin of the measured $\gamma$-ray emission would 
be favored (K13). 

The dataset used in our analysis 
is specified in Table\,2, where the listed fluxes refer to the combined lobes. We take 0.03-5\,GHz radio fluxes from 
Jones et al. (2012; and references therein), 30 GHz {\it Planck} and 0.2-20\,GeV {\it Fermi}-LAT data from K13, and 
the $1$\,keV {\it ASCA} flux from Tashiro et al. (1998). Concerning X-ray emission, we assume the 2-10\,keV {\it Suzaku} 
upper limit to be consistent with the {\it ASCA} flux (see K13), so use the latter 1\,keV flux.

\begin{table*}
\caption[] {Cen\,B: lobe emission.}
\begin{tabular}{ c  c  l  l  c  c  l}
\hline
\hline
\noalign{\smallskip}
Frequency& Flux density$^{\star}$&   Reference                  &   &Frequency         & Flux density$^{\star}$ & Reference            \\
log($\nu$/Hz)&                   &                              &   &log($\nu$/Hz)     &                        &                      \\
\noalign{\smallskip}
\hline
\noalign{\smallskip}
$7.477$ &  $1684 \pm 118$ Jy & Finlay \& Jones (1973)           &   & $~\,9.423$       & $60 \pm 6$ Jy            & Milne \& Hill (1969) \\
$7.932$ &  $795$ Jy          & Mills, Slee \& Hill (1971)       &   & $~\,9.431$       & $61$       Jy            & Day, Thomas \& Goss (1969) \\
$8.004$ &  $750$ Jy          & Mills (1952)                     &   & $~\,9.699$       & $39$       Jy            & Goss \& Shaver (1970)\\
$8.611$ &  $242$ Jy          & Komessaroff (1966)               &   & $10.477$         & $12 \pm 1$ Jy            & Katsuta et al. 2013  \\
$8.611$ &  $210$ Jy          & Kesteven (1968)                  &   & $17.383$         & $0.22 \pm 0.09$ $\mu$Jy  & Tashiro et al. 1998  \\
$8.611$ &  $136$ Jy          & Shaver \& Goss (1970)            &   & $22.94 \pm 0.25$ & $6.5 \pm 2.5$ pJy        & Katsuta et al. 2013  \\
$8.926$ &  $150$ Jy          & McAdam (1991)                    &   & $23.43 \pm 0.25$ & $1.1 \pm 0.5$ pJy        & Katsuta et al. 2013  \\
$9.149$ &  $102 \pm 10$ Jy   & Milne \& Hill (1969)             &   & $23.94 \pm 0.25$ & $0.24 \pm 0.10$ pJy      & Katsuta et al. 2013  \\ 
$9.158$ &  $130$ Jy          & Mathewson, Healey \& Rome (1962) &   & $24.43 \pm 0.25$ & $<0.11$ pJy              & Katsuta et al. 2013  \\ 
\noalign{\smallskip}
\hline
\hline
\end{tabular}
\begin{flushleft}
\noindent
$^{\star}$Where not explicitly indicated, flux density uncertainties are assumed to be $10\%$. 
\end{flushleft}
\end{table*}

\subsection{NGC\,6251}

\begin{figure}
\vspace{12.5cm}
\includegraphics{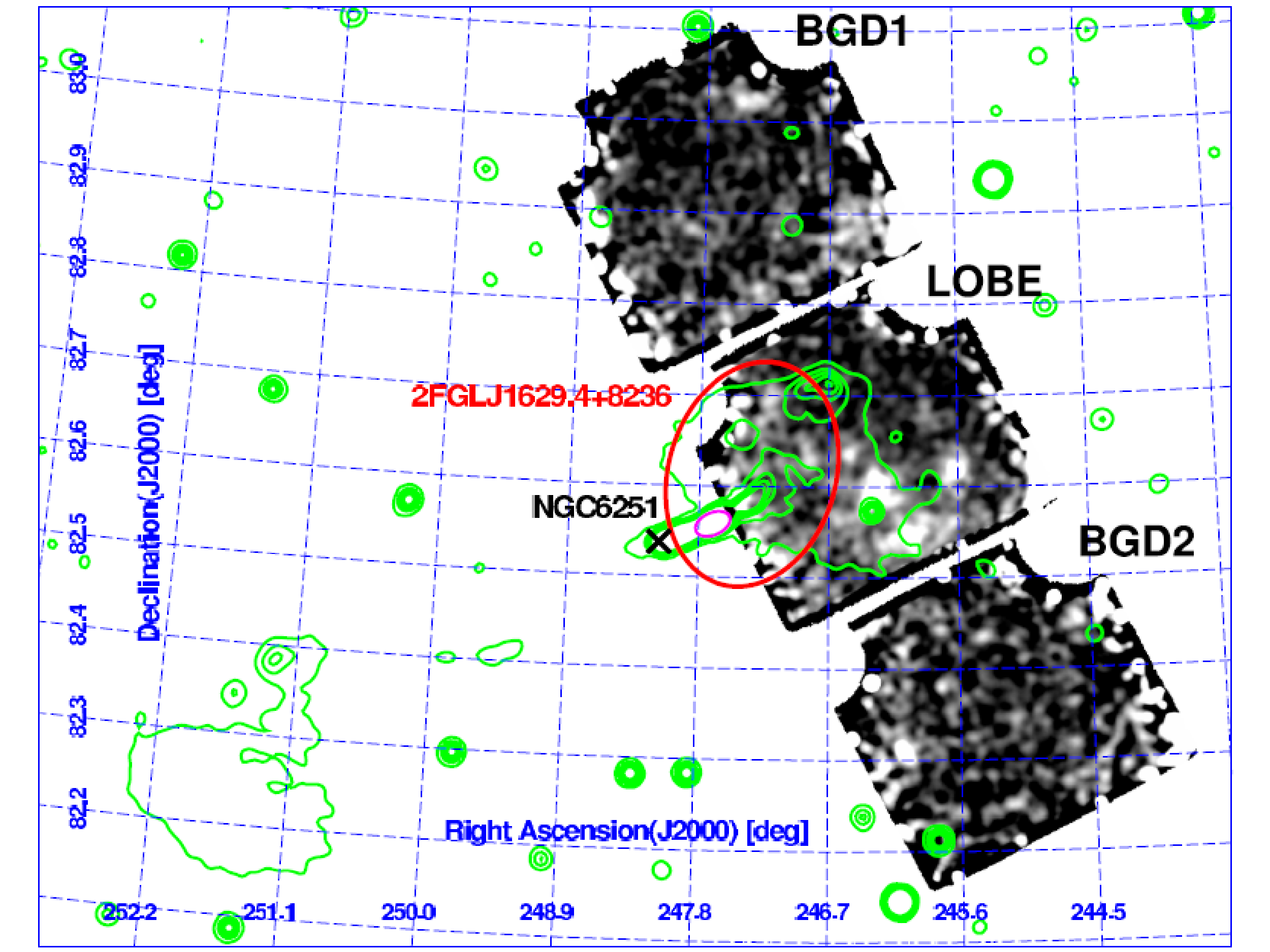}
\includegraphics{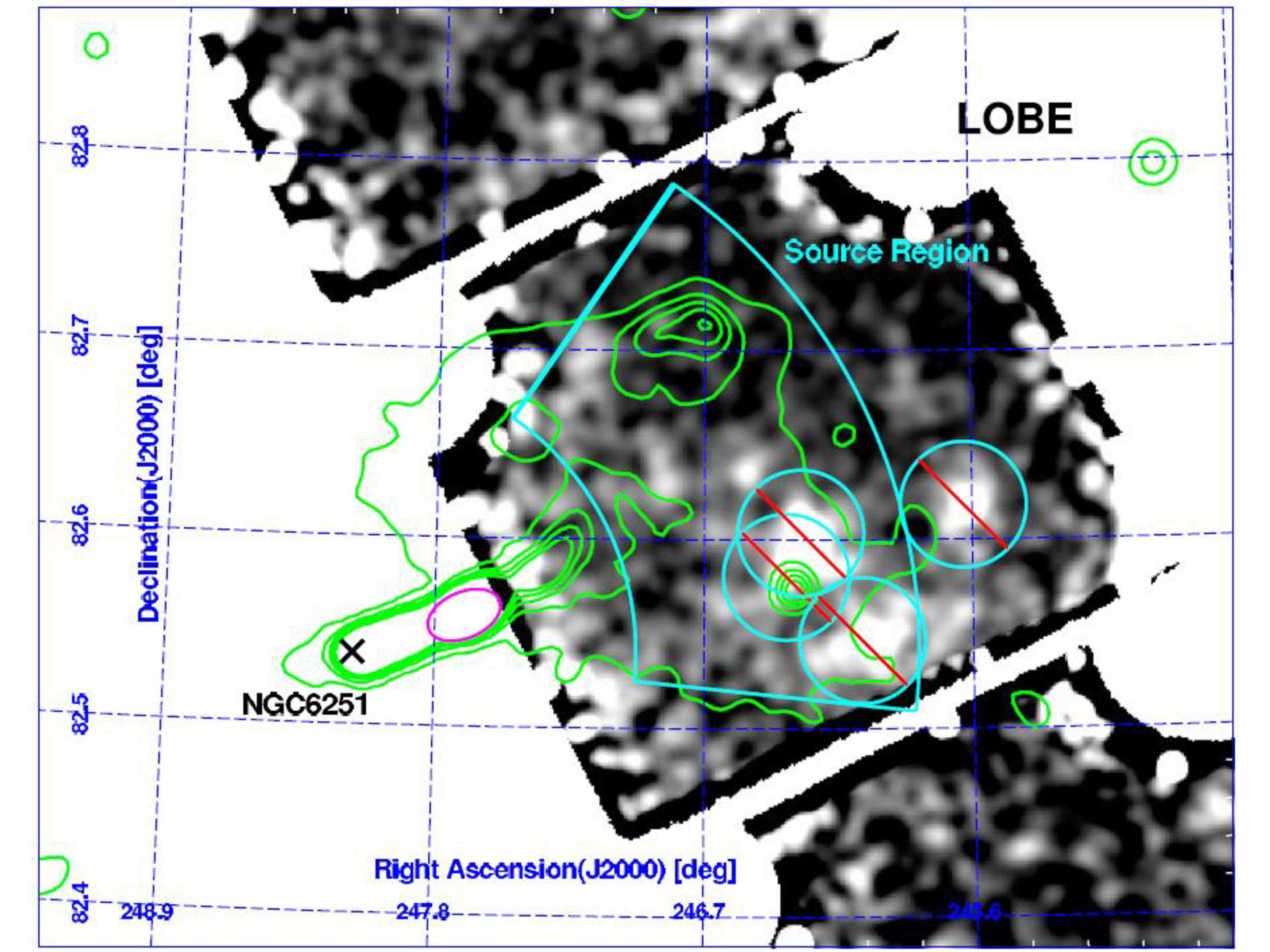}
\caption{
Image of the NW lobe region of NGC\,6251 (reproduced with permission from Figures\,1 and 2 of T12). The image shows the 
relative excess of the smoothed {\it Suzaku} X-ray (0.4-10 keV) photon counts (see T12 for details). 
{\it Left.} The radio contours (green in the online version) denote the large-scale structure observed by the Westerbork 
Synthesis Radio Telescope at $55^{\prime \prime}$ resolution (Mack et al. 1997) and indicate levels of 8, 31, 54, 77, and 
100 mJy/beam. The ellipse (red in the online version) denotes the 95\% position error of 2FGL\,J1629.4+8236. The position 
of the radio core of NGC\,6251 is marked by the cross at the center, and the adjacent outer jet region is marked by the 
ellipse (magenta in the online version).
{\it Bottom.} Same as top panel, but zooming into the NW lobe region. The Suzaku source extraction region is denoted by 
the section of circular corona (cyan in the online version). The point X-ray sources removed from the T12 analysis are 
marked by the circles with stripes (cyan and red, respectively, in the online version). 
}
\label{fig:N6251}
\end{figure}

Located at at $D_L = 106$\,Mpc, NGC\,6251 has the largest angular extent among northern radio sources, $1.2^o$. The 
NW radio lobe was detected by {\it Suzaku} to be very extended, likely associated with {\it Fermi}-LAT source 
2FGL\,J1629.4+8326 (Takeuchi et al. 2012, hereafter T12; see Fig.\,\ref{fig:N6251}). 
The observed $\gamma$-ray emission was interpreted as Compton scattering of the radio-emitting electrons off the CMB 
(T12). 

The dataset used in our analysis (Table\,3) comes from T12: the 0.327-1.56\,GHz Westerbork Synthesis Radio Telescope 
(WSRT) and Very Large Array fluxes are measured using the same source extraction regions as in the 0.5-8\,keV {\it Suzaku} 
analysis.


\begin{table*}
\caption[] {NGC\,6251: NW lobe emission.}

\begin{tabular}{ c  c  l  l  c  c  l}
\hline
\hline

\noalign{\smallskip}
Frequency$^{\star}$& Flux density & Reference             &   &Frequency$^{\star}$&  Flux density         & Reference            \\
log($\nu$/Hz)     &               &                       &   &log($\nu$/Hz)      &                       &                      \\
\noalign{\smallskip}
\hline
\noalign{\smallskip}
$~\,8.515$ & $3.10 \pm 0.29$ Jy  & Takeuchi et al. (2012) &   & $17.860$          & $25 \pm 4$ nJy        & Takeuchi et al. (2012) \\
$~\,8.785$ & $1.75 \pm 0.17$ Jy  & Takeuchi et al. (2012) &   & $18.161$          & $12 \pm 5$ nJy        & Takeuchi et al. (2012) \\
$~\,9.193$ & $0.91 \pm 0.09$ Jy  & Takeuchi et al. (2012) &   & $22.542 \pm 0.25$ & $7.2 \pm 2.6$ pJy     & Takeuchi et al. (2012) \\
$10.023$   & $0.21 \pm 0.03$ Jy  & Mack, Klein, O'Dea \& Willis (1997)     &   & $23.034 \pm 0.25$ & $3.2 \pm 1.9$ pJy     & Takeuchi et al. (2012) \\
$17.258$   & $44 \pm 12$ nJy     & Takeuchi et al. (2012) &   & $23.542 \pm 0.25$ & $0.63 \pm 0.19$ pJy   & Takeuchi et al. (2012) \\
$17.559$   & $46 \pm 10$ nJy     & Takeuchi et al. (2012) &   & $24.034 \pm 0.25$ & $0.083 \pm 0.024$ pJy & Takeuchi et al. (2012) \\ 

\noalign{\smallskip}

\hline
\hline

\end{tabular}
\begin{flushleft}
\noindent
$^{\star}$The X-ray flux densities refer to the photon energy intervals 0.5-1\,keV, 1-2\,keV, 2-4\,keV, 4-8\,keV.
\end{flushleft}

\end{table*}


\section{Radiation fields in the lobes}

A reasonably precise determination of the ambient photon fields in the lobes is needed for predicting the 
level of $\gamma$-ray emission from Compton scattering of the radio-emitting electrons (and positrons). 
Due to the proximity of the central active galaxy, radiation fields in the lobes include local, in addition 
to cosmic (background) components.

Relevant cosmic radiation fields include the CMB and the EBL. The CMB is a pure Planckian with $T_{\rm CMB} 
= 2.735$\,K and energy density $u_{\rm CMB} = 0.25\,(1+z)^4$ eV cm$^{-3}$ (e.g. Dermer \& Menon 2009). 
The EBL originates from direct and dust-reprocessed starlight integrated over the star formation history of 
the universe. A recent updated model, based on accurate galaxy counts in several spectral bands, is presented 
by Franceschini \& Rodighiero (2017, 2018); its main components are the cosmic IR background (CIB) and the 
cosmic optical background (COB), described as diluted Planckians (PR19). We approximate the EBL as a combination 
of diluted Planckians, 
\begin{eqnarray}
\lefteqn{
n_{\rm EBL}(\epsilon) ~=~ \sum_{j=1}^7 A_j \,{8 \pi \over h^3c^3} \, \frac{\epsilon^2}{e^{\epsilon/k_B T_j}-1} \hspace{0.5cm}  {\rm cm^{-3}~ erg^{-1}} }
\label{eq:EBL}
\end{eqnarray}
with 
$A_1=10^{-5.629}$, $T_1=29$\,K, 
$A_2=10^{-8.496}$, $T_2=96.7$\,K, 
$A_3=10^{-10.249}$, $T_3=223$\,K, 
$A_4=10^{-12.027}$, $T_4=580$\,K,  
$A_5=10^{-13.726}$, $T_5=2900$\,K, 
$A_6=10^{-15.027}$, $T_6=4350$\,K, 
$A_7=10^{-16.364}$, $T_7=8700$\,K,
that incorporates galaxy-counts based results (Franceschini et al. 2008; Franceschini \& Rodighiero 2017, 2018) 
and $\gamma$-ray derived results (that suggest a slight enhancement of the optical hump; Acciari et al. 2019).

Local radiation fields, that constitute the Galactic Foregound Light (GFL), arise from the central 
elliptical galaxies located between the lobes. Galaxy SEDs usually show two thermal humps, IR and 
optical.
\smallskip

\noindent
{\it Cen\,A}:  The central galaxy, NGC\,5128, has a bolometric IR luminosity of $L_{\rm IR} \simeq 4.8 
\cdot 10^{43}$\,erg\,s$^{-1}$,  as implied by {\it IRAS} flux densities at 12, 25, 60 and 100$\mu$m (Gil 
de Paz et al. 2007)
\footnote{
The total IR ($8-1000\,m$m) flux is $f_{\rm IR}=1.8 \cdot 10^{-11} (13.48\,f_{12} + 5.16\,f_{25} + 
2.58\,f_{60} + f_{100})$\, erg\,cm$^{-2}$s$^{-1}$ (Sanders \& Mirabel 1996). }
. The optical bolometric luminosity, $L_{\rm opt} \sim 1.6 \cdot 10^{44}$\,erg\,s$^{-1}$, is derived 
from the total $B$-band magnitude (Gil de Paz et al. 2007), converted to bolometric magnitude using 
$B-V \sim 0.9$ (Dufour et al. 1979) following the procedure outlined in PR19. The surface brightness 
distribution extends out to $\sim 625^{\prime \prime}$ following a $(R/R_e)^{1/4}$ profile with $R_e = 
305^{\prime \prime}$ (=11.5\,kpc; Dufour et al. 1979). This effective radius is modest as compared with 
the distance ($70$\,kpc) to the nearest boundaries of the innermost lobe regions N3 and S3. 
\smallskip

\noindent
{\it Cen\,B}: The host galaxy PKS\,1343-601, which is projected close to the Galactic plane, is extremely 
absorbed ($A_B \sim 12.3$ mag, Schr\"oder et al. 2007). We estimate $L_{\rm opt} \sim 10^{45}$\,erg\,s$^{-1}$ 
from $B \sim 11.8$ mag (total, extinction corrected; Schr\"oder et al. 2007) converted to bolometric magnitude 
assuming $B-V = 0.95$ (a typical value for ellipticals) and following the procedure outlined in PR19. Also, 
$L_{\rm IR} \sim 4 \cdot 10^{44}$ erg\,s$^{-1}$ from the B-luminosity through a FIR--B scaling relation 
(Bregman et al. 1998) and setting $L_{\rm IR} \sim 2\,L_{\rm FIR}$ (e.g. Persic \& Rephaeli 2007). The surface 
brightness distribution is unknown; however, if it has a typical $R^{1/4}$ profile, its effective radius can be 
estimated from the B-luminosity (Romanishin 1986): $R_e \sim 17$\,kpc. Lobe radii and nearest lobe boundaries 
are $\sim 100$\,kpc (Jones et al. 2001, K13). 
\smallskip

\noindent
{\it NGC\,6251}: We estimate $L_{\rm opt} \sim 10^{45}$ erg\,s$^{-1}$ from $B_T^0 = 13.22$ (RC3 catalogue), and 
following the steps outlined in the previous case, $L_{\rm IR} \sim 4 \cdot 10^{43}$\,erg\,s$^{-1}$ from {\it IRAS} 
flux densities (Golombek et al. 1988). The relevant surface brightness distribution is a modified Hubble profile 
($\propto [1+(r/a)^2] ^{-1}$) with $a=1.1^{\prime\prime}$ ($0.565$\,kpc; Crane et al. 1993). The lobe radius and 
nearest boundary distance are $185$\,kpc and $265$\,kpc (see T12).
\smallskip

The above IR and optical parameters allow modeling the GFL. In our calculations we take $T_{\rm gal,\,OPT} = 2900$\,K 
and $T_{\rm gal,\,IR} = 29$\,K (see PR19). The $\gamma$-ray data for the lobe regions of Cen\,A, and the lobes of 
Cen\,B and NGC\,6251 are spatial averages. Similarly, we compute the corresponding volume-averaged Compton/GFL yield. 
As the characteristic radii of NGC\,5128, PKS\,1343-601, NGC\,6251 are small compared with those of the corresponding 
lobes (radius, distance from central galaxy), we treat central galaxies as point sources
\footnote{ 
Assuming the lobe is a sphere with radius $r_s$, centered at $C(d+r_s,\,0)$, with $d$ 
the $x$-coordinate of the nearest lobe boundary. A vertical line $x=d+u$ (being $u$ the 
$x$-axis running variable inside the lobe, $0 \leq u \leq 2 r_s$) intersects the lobe 
at $y_t = \pm \sqrt{ -(d+u)^2 + 2 (d+r_s) (d+u) - d (d+2r_s) }$. The photon energy 
density at the lobe boundary is $u_{\rm gal}(d) = L_{\rm opt}/(4 \pi d^2 c)$. Since 
the lobe is symmetric w.r.t. the $x$-axis, the volume-averaged 
%
%
energy density is $\bar u_{\rm gal} = \psi \, u_{\rm opt}(d)$, where $\psi ~=~ V_{\rm lobe}^{-1} \int_0^{2r_s} \int_0^{y_t} 
{d^2 \over (d+u)^2+y^2} ~ 2\pi y~ dy ~ du$ with $V_{\rm lobe} = (4/3) \pi r_s^3$. This formula applies also to a cylindrical 
lobe (radius $r_s$, height $r_h$, and base located at a distance $d$ from the central galaxy on the $x$-axis),  
but with $y_t = r_s$ and $V_{\rm lobe} = \pi r_s^2 r_h$. 
}
(in contrast with our procedure for Fornax\,A in PR19).

\section{Modeling NT emission}

\subsection{Synchrotron and Compton yields}

Radio emission in the lobes is by electron synchrotron in a disordered magnetic field whose mean 
value $B$ is taken to be spatially uniform. X-$\gamma$ emission is modeled to originate in Compton 
scattering of the electrons off the CMB and optical radiation fields. The standard emissivity 
calculations are briefly described in PR19.

Assuming no appreciable temporal flux variation, the electron energy distribution (EED) is taken to be 
time-independent, spatially isotropic and truncated-PL distribution in the electron Lorentz factor, 
$N_e(\gamma) = N_{e,0}\, \gamma^{-q_e}$ in the interval $[\gamma_{min},\, \gamma_{max}$], with 
a finite $\gamma_{max}$. The electrons traverse magnetized lobe plasma, with field strength $B$, and 
scatter off the ambient radiation fields emitting radio synchrotron and Compton X-and-$\gamma$-ray 
radiation (see Eqs.\,9 and 11 of PR19). The results of our lobe SED modeling are summarized below 
and shown in Figs.\,1 and 2.
\smallskip

\noindent
{\it Cen\,A}. 
Measured NT emission at $1$\,keV, pivotal to our previous analysis of Fornax\,A (PR19), is unavailable; diffuse 
emission from a lobe region of Cen\,A measured by {\it XMM-Newton} was determined to be of thermal origin (Kraft 
et al. 2003). Thus, to calibrate the EED in each lobe region we either maximized the Compton/CMB contribution to 
the lowest-frequency {\it Fermi}-LAT point, or normalized it to the {\it Fermi}-LAT data, if the spectral shape 
of the the data matched that of the predicted Compton/EBL spectrum. Fitting the predicted synchrotron flux to 
radio data yielded the EED spectral index, $q_e$. The upper energy cutoff, $\gamma_{\max}$, was determined by 
requiring the predicted synchrotron flux to fit the radio SED turnover (e.g., region N1) or, if the latter is not 
clearly suggested by the data, by requiring the predicted Compton/EBL flux not to overproduce the lowest-energy 
{\it Fermi}-LAT data point. The minimum electron energy, $\gamma_{\rm min}$, cannot be derived directly from the 
present data set owing to the lack of low-energy (1\,keV) Compton/CMB data. We can only estimate $\gamma_{\rm 
min}$ by assuming it marks the transition from the Coulomb-loss regime to the synchrotron/Compton-loss regime, 
$\gamma_{\rm t} \simeq 10^2 (\frac{n_{\rm gas}}{10^{-4}\, {\rm cm}^{-3}})^{1/2} (\frac{u_B}{0.025\,{\rm eV\, 
cm}^{-3}} + \frac{u_r}{0.25\,{\rm eV\, cm}^{-3}})^{-1/2}$ (e.g. Rephaeli \& Persic 2015) -- where the 
magnetic and radiation energy densities, $u_B$ and $u_r$, are scaled to $B=1\,\mu$G and the CMB radiation field, 
respectively. Doing so yielded $\gamma_{\rm min} \sim 100$. The mean magnetic field strength, $B$, was deduced 
by adjusting the predicted synchrotron yield to the radio measurements. 

\begin{figure*}
\vspace{12.5cm}
\includegraphics{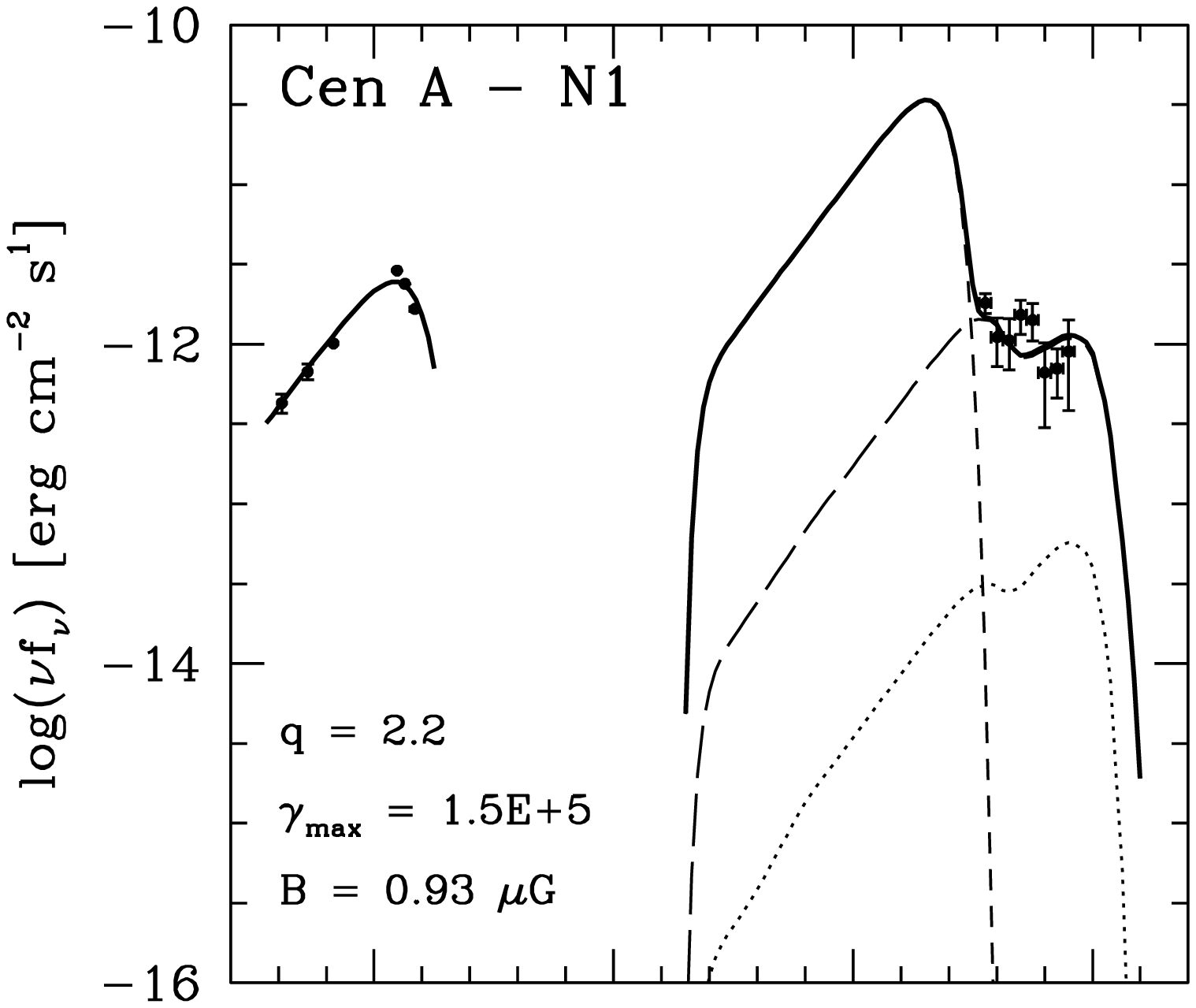}
\includegraphics{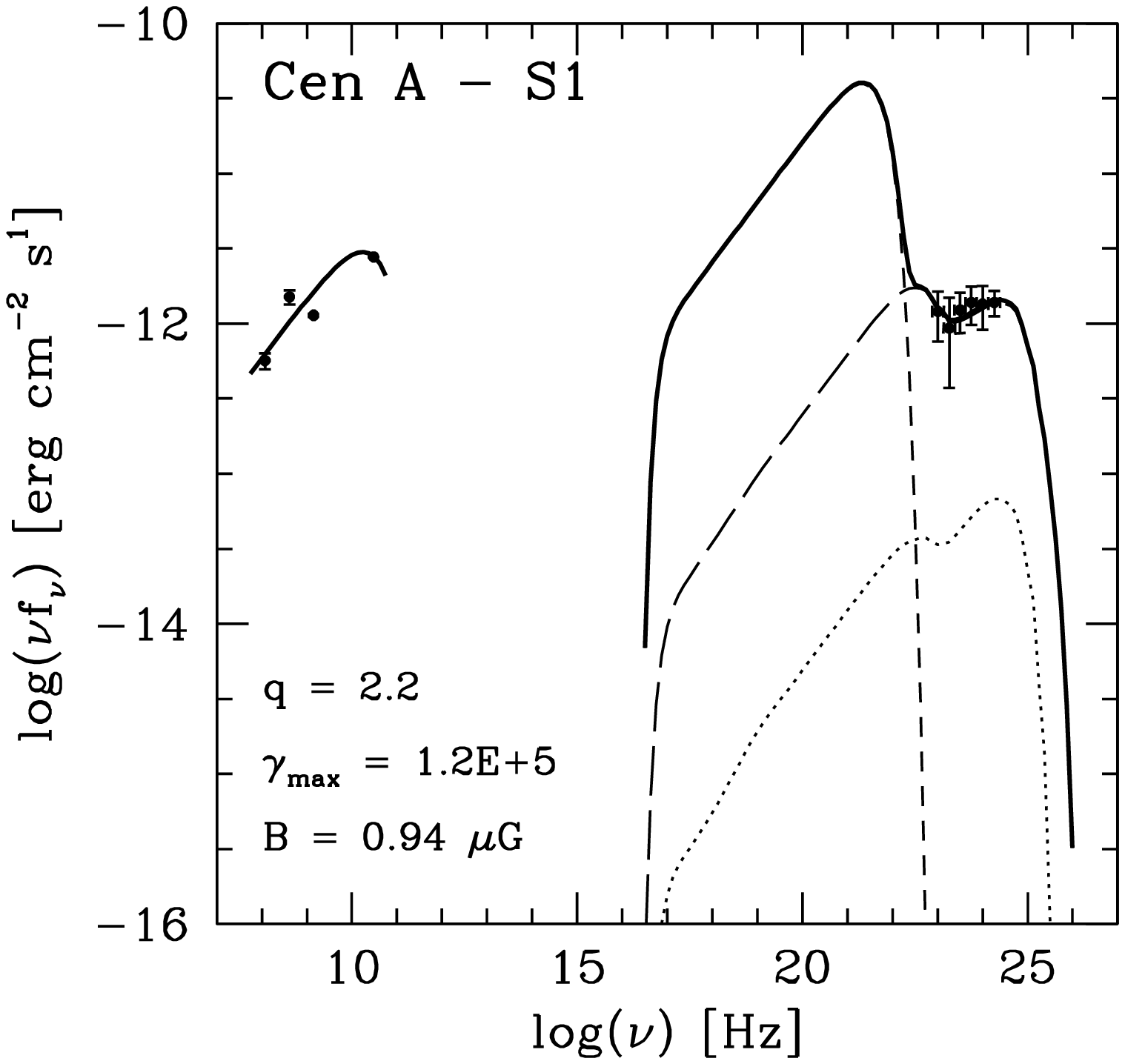}
\includegraphics{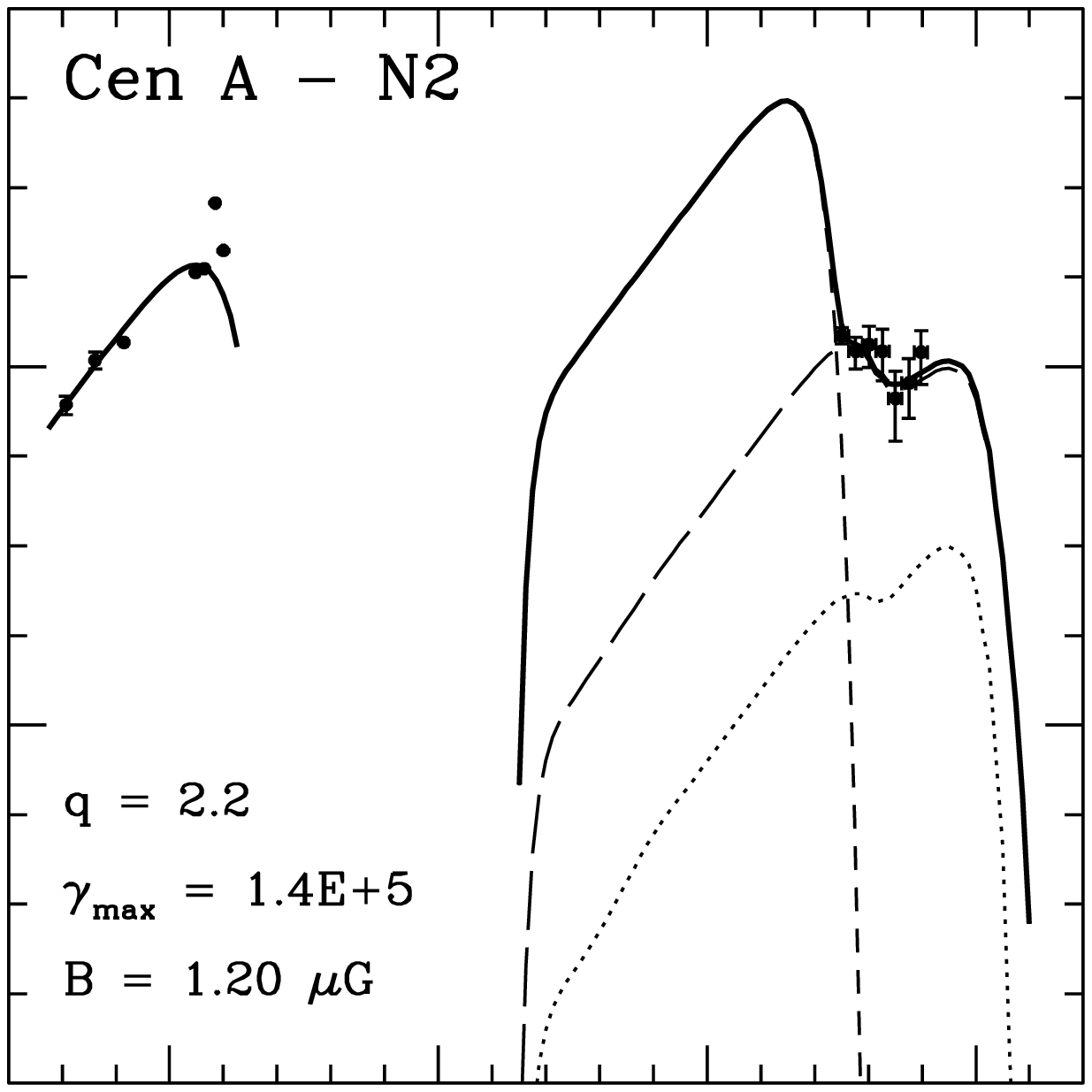}
\includegraphics{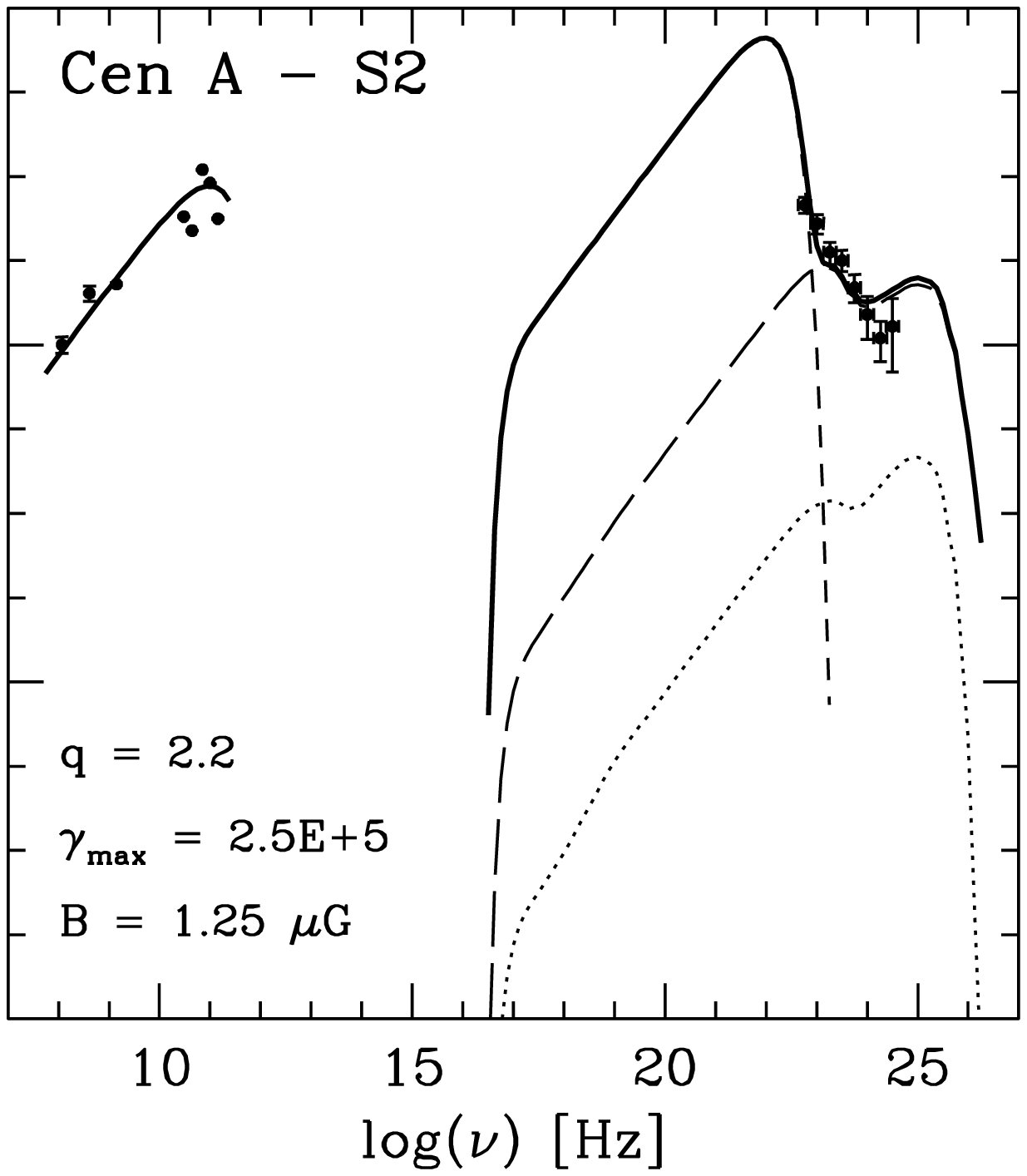}
\includegraphics{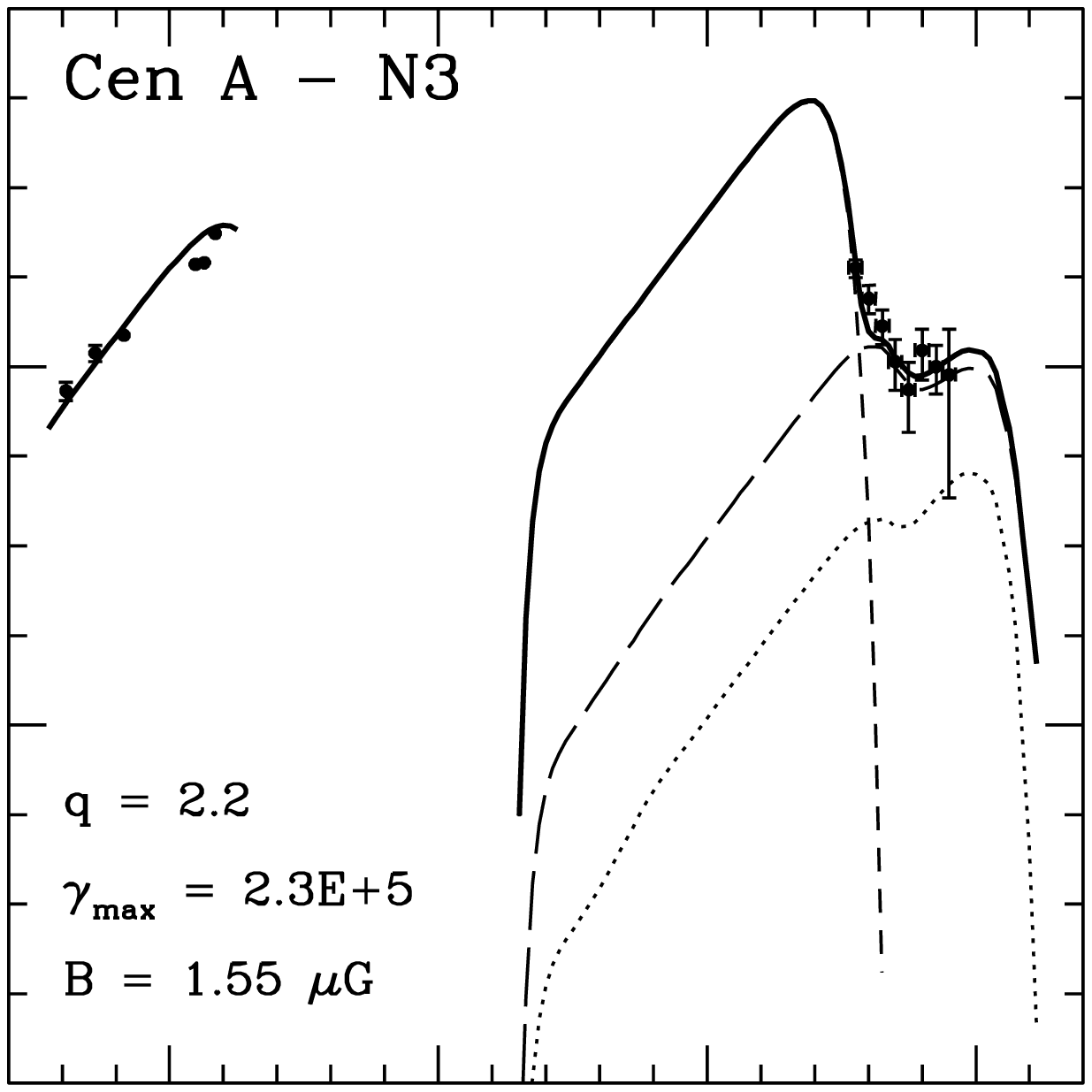}
\includegraphics{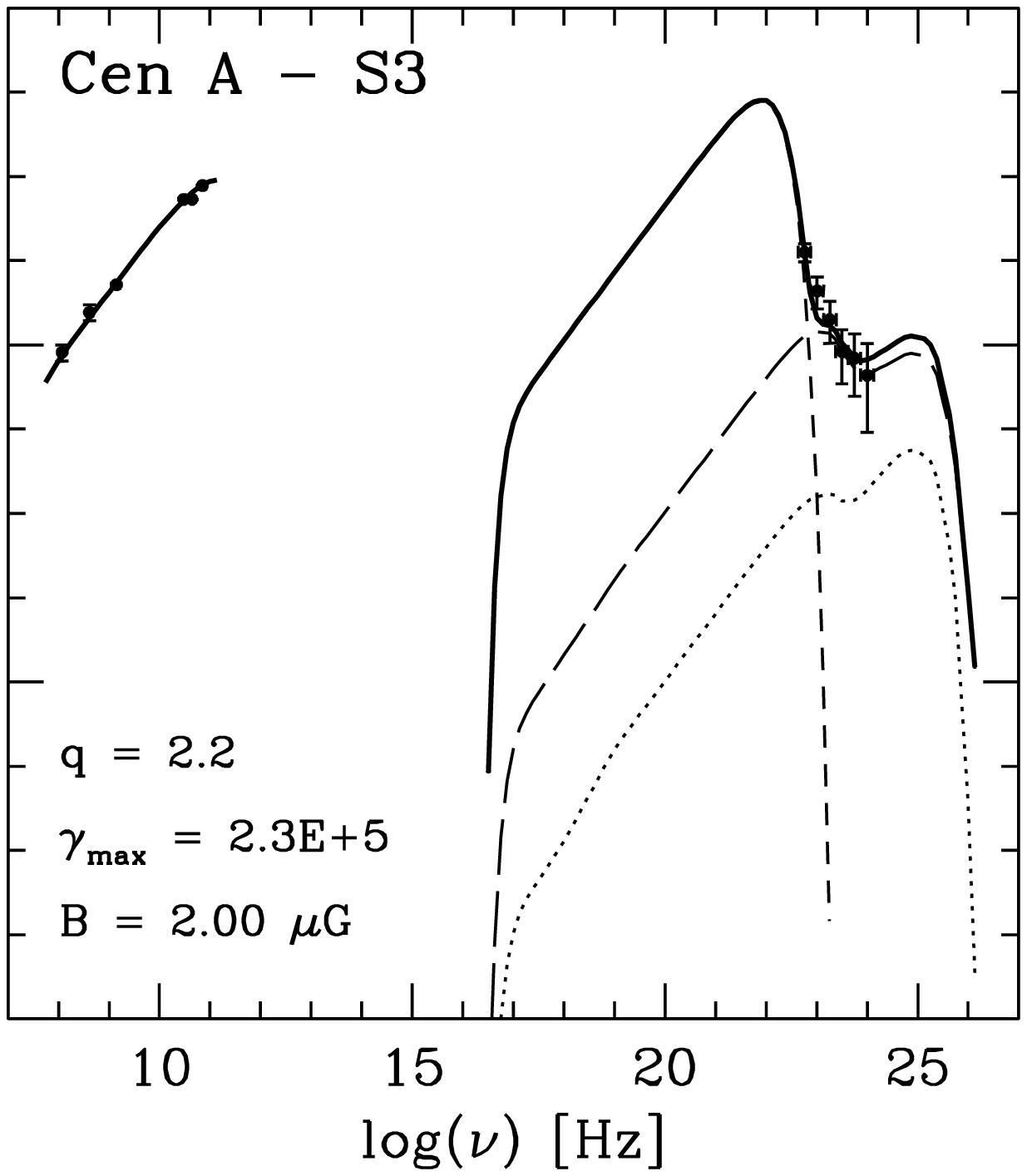}
\caption{
Predicted and measured SEDs for the six lobe regions of Cen\,A specified in SYMA16. Regions N1, S1 are 
farthest from the host galaxy NGC\,5128; regions N2, S2 are intermediate; regions N3, S3 are closest. 
Data are denoted by dots (with error bars). Emission component curves are: synchrotron, solid; Compton/CMB, 
short-dashed; Compton/EBL, long-dashed; Compton/GFL, dotted; total Compton: thick solid. Indicated in each 
panel are the values of the EED parameters. 
}
\label{fig:CenA_SED}
\end{figure*}

The results are shown in Fig.\,\ref{fig:CenA_SED} together with the radio and $\gamma$-ray measurements. 
In all lobe regions the predicted Compton emission is consistent with the $\sim$0.1-10\,GeV {\it Fermi}-LAT 
data. The dominant radiation field is the EBL, as has been previously suggested, using different EEDs and EBL 
models (Abdo et al. 2010, SYMA16). The maximum electron energy, $\gamma_{\rm max}$, is lower by a factor $\sim 
2$ in the outer as compared to the inner regions (see Fig.\,\ref{fig:CenA_SED}). The less energetic EEDs 
necessarily sample more energetic photons from the target radiation fields to produce Compton emission in the 
{\it Fermi}-LAT band. Indeed, a virtually pure EBL shape is revealed by the {\it Fermi}-LAT data in the outer 
regions, whereas a significant contribution from the CMB profile is recognizable in the inner regions. 
\smallskip

\begin{figure*}
\vspace{7.0cm}
\includegraphics{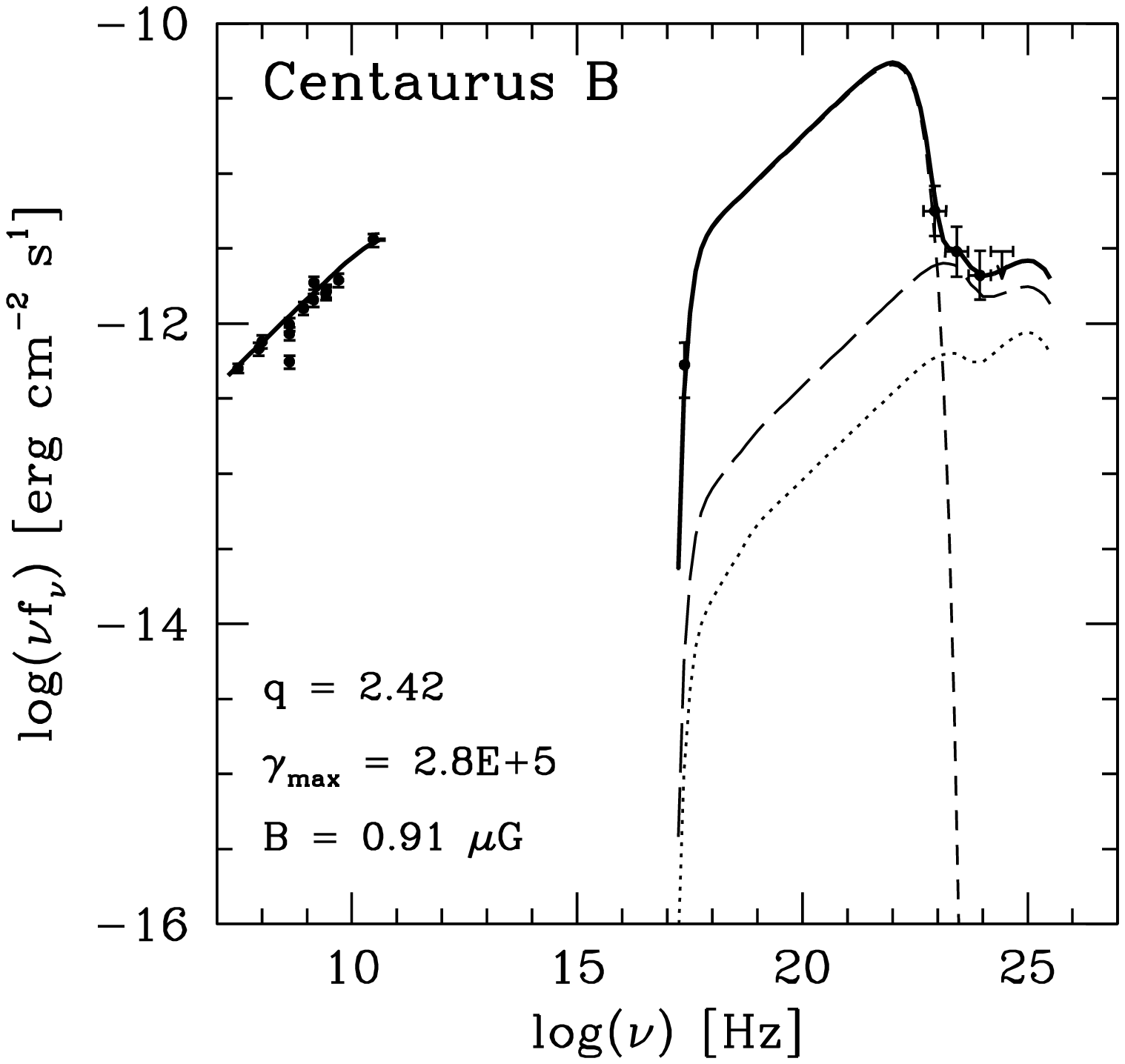}
\includegraphics{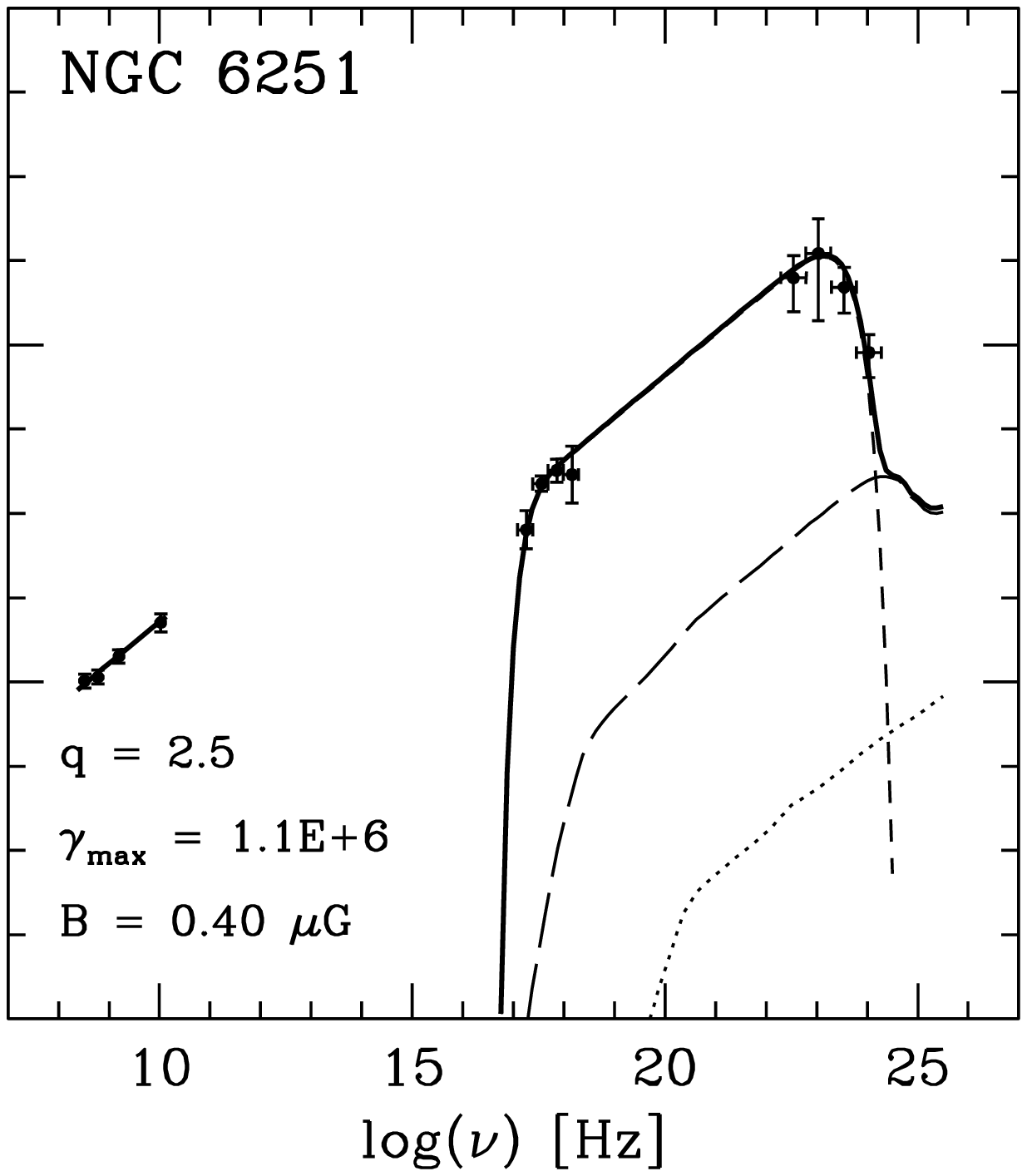}
\includegraphics{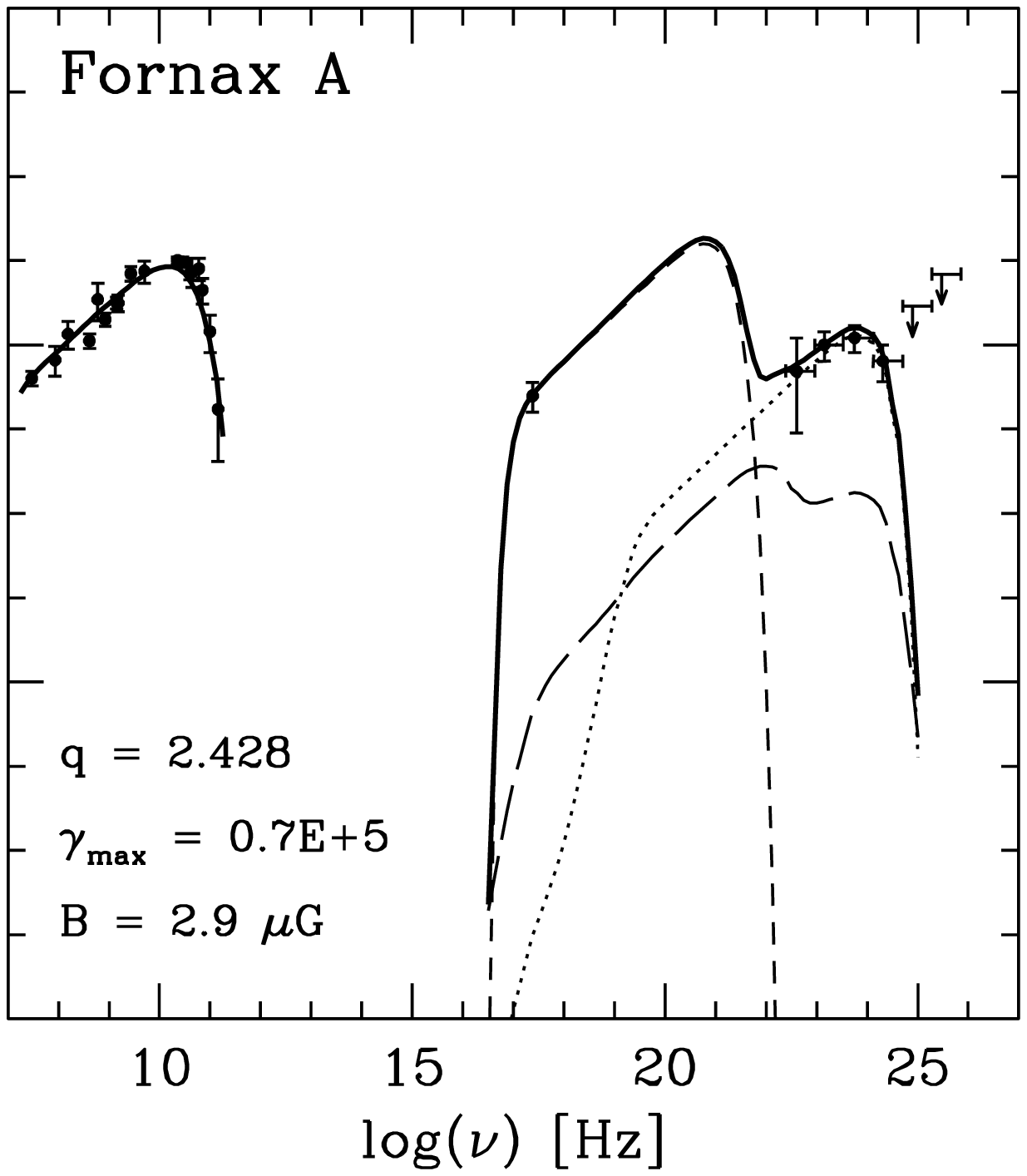}
\caption{ 
Broad-band SEDs of Cen\,B and NGC\,6251. For comparison, the SED of Fornax\,A is also shown: the 
model is as in PR19 except for implementing the current EBL model and neglecting secondary yields. 
Emissions and model parameters are as in Fig.\,\ref{fig:CenA_SED}.
}
\label{fig:CenB_N6251_FornaxA_SED}
\end{figure*}

\noindent
{\it Cen\,B}.
We use the photoelectrically absorbed ($N_H = 1.06 \cdot 10^{22}$ cm$^2$; K13) 1\,keV flux density to determine 
$N_{e,0}$ assuming the emission is Compton/CMB. Spectral fitting the predicted spectrum to the radio data yields 
$q_e$, whereas $\gamma_{\rm max}$ is deduced by requiring the predicted Compton/CMB flux not to exceed the 
lowest-energy {\it Fermi}-LAT data point; we set $\gamma_{\rm min}=100$. With the electron spectrum fully specified, 
normalization of the predicted synchrotron spectral flux to the radio measurements yields $B$. The SED is shown 
in Fig.\,\ref{fig:CenB_N6251_FornaxA_SED}-left. {\it Fermi}-LAT data are modeled as Compton scattering off the CMB 
at lower energies and as Compton scattering off the EBL (and subdominant GFL) at higher energies. The $\gamma$-ray 
spectrum clearly reflects the shape of the Compton/EBL(+GFL). 
\smallskip

\noindent
{\it NGC\,6251}.
The spectral index $q_e$ is deduced as described above. Modeling the $\gamma$-ray spectrum and the photoelectrically 
absorbed ($N_H = 5.54 \cdot 10^{20}$ cm$^2$, Dickey \& Lockman 1990) X-ray flux as Compton/CMB yields $\gamma_{\rm 
max}$, and $N_{e,0}$ and $\gamma_{\rm min}=600$. Adjusting the predicted synchrotron yield to the radio data determines 
$B$. The SED is shown in Fig.\,\ref{fig:CenB_N6251_FornaxA_SED}-middle; the X-ray and $\gamma$-ray data are interpreted as 
Compton/CMB emission, with the predicted Compton/EBL flux amounting to just few percent of the {\it Fermi}-LAT flux 
(and negligible contribution from Compton/GFL).  
\smallskip

%
%

As is apparent from Figs.\,\ref{fig:CenA_SED},\ref{fig:CenB_N6251_FornaxA_SED}, the radiative yields of energetic 
electrons in the lobes of Cen\,A, Cen\,B, and NGC\,6251 can adequately account for all currently available radio, X-ray, 
and $\gamma$-ray measurements. This basic result of our spectral analyses constitutes further substantiation of what has 
been suggested in previous works (SYMA16, Yang et al. 2012; K13; T12). Note also that GFL contributes insignificantly to 
the measured levels of $\gamma$-ray emission, 
\footnote{
We use projected galaxy-to-lobe distances (inclinations are unknown), 
so inferred GFL densities are strict upper limits.
}
in clear contrast with our finding in the case of Fornax\,A, in whose lobes the main component of $\gamma$-ray 
emission is by Compton scattering off GFL (PR19). The double-humped profile in the predicted $\gamma$-ray spectra 
of Cen\,A and Cen\,B (Figs.\,1 and 2) mimics the spectral distribution of the EBL (and GFL); this underlines its 
origin. Also, the sharp $\gamma$-ray cutoff in the spectrum of NGC\,6251 reflects the truncation of the electron 
energy distribution.Thus, these features by themselves indicate that the $\gamma$-ray spectral profiles of these 
sources are unlikely to be of hadronic origin that has the characteristic hump-shaped pion-decay profile. It seems 
that this {\it a priori} rules out any significant pionic component to these SEDs. (In contrast, the $\gamma$-ray 
profile of Fornax\,A is amenable to a pionic interpretation; McKinley et al. 2015, Ackermann et al. 2016.)

\subsection{Upper limits on proton contents in lobes: \\
the case of Cen\,A}

Energetic particles in the lobes are thought to be injected by jets emanating from the central galaxy. There 
is mounting evidence (e.g. Krawczynski et al. 2012) that AGN jets are energetically and dynamically dominated 
by protons, albeit with an uncertain proton-to-electron number density ratio. This is deduced from limits on 
the electron (and positron) contents of jets deduced from the lack of soft-X-ray excesses, and based on insight 
gained from leptonic models of knots and blazars. Furthermore, X-ray cavities in galaxy clusters (interpreted 
as expanding bubbles from radio sources) require a significant pressure component, in addition to that of 
energetic electrons and magnetic fields, in order to overcome thermal intracluster gas pressure (e.g. Wilson et 
al. 2006). Jet protons can undergo photopion production through {\it p--}$\gamma$ interactions, and most of the 
proton energy is eventually channeled into pair production. Due to their less efficient radiative losses, protons 
can be accelerated to much higher final energies than electrons. Acceleration is likely distributed along the jet 
because adiabatic and radiative cooling would cause the jet surface brightness to decrease much faster with 
galactocentric radius than observed (Blandford et al. 2018). If so, the electrically neutral NT plasma injected 
by the jet into the lobes would have a similar spectral index for protons and electrons, so their energy 
densities may be related, $u_p/u_e \sim (m_p/m_e)^{(3-q)/2}$ (e.g. Persic \& Rephaeli 2014). 

The pionic yield in the lobes can be estimated following the formalism in PR19. Assuming the proton energy 
distribution (PED) to be $N(E) = N_{p0} E^{-q_p}$ (with $E$ in GeV) in the interval 1-500 GeV and the thermal 
plasma to have a nominal density $n_{\rm gas} = 10^{-4}$ cm$^{-3}$ (Stawarz et al. 2013 for Cen\,A), an upper 
limit (UL) to the proton content can be deduced from spectral considerations.

\begin{figure*}
\vspace{12.5cm}
\includegraphics{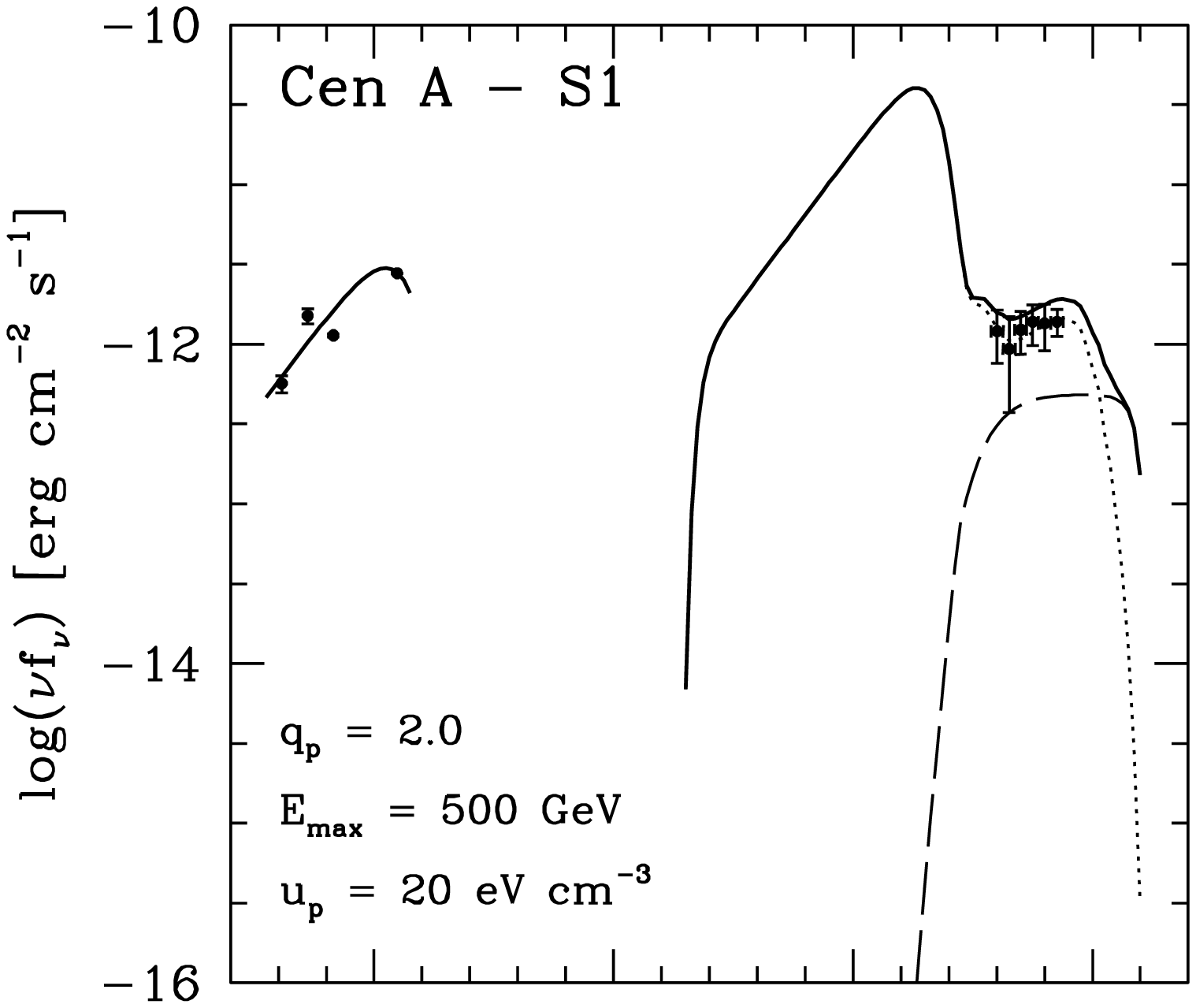}
\includegraphics{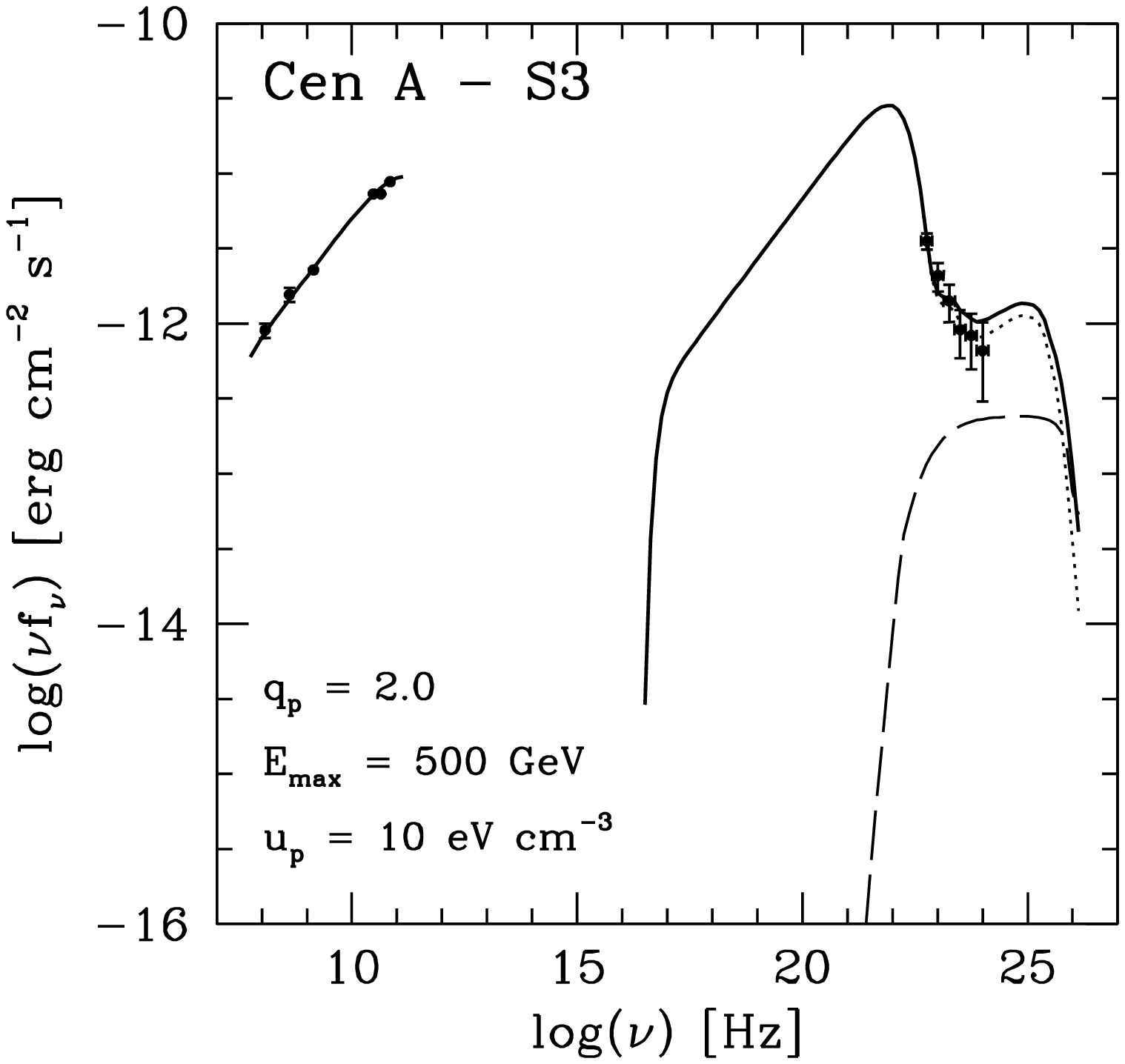}
\includegraphics{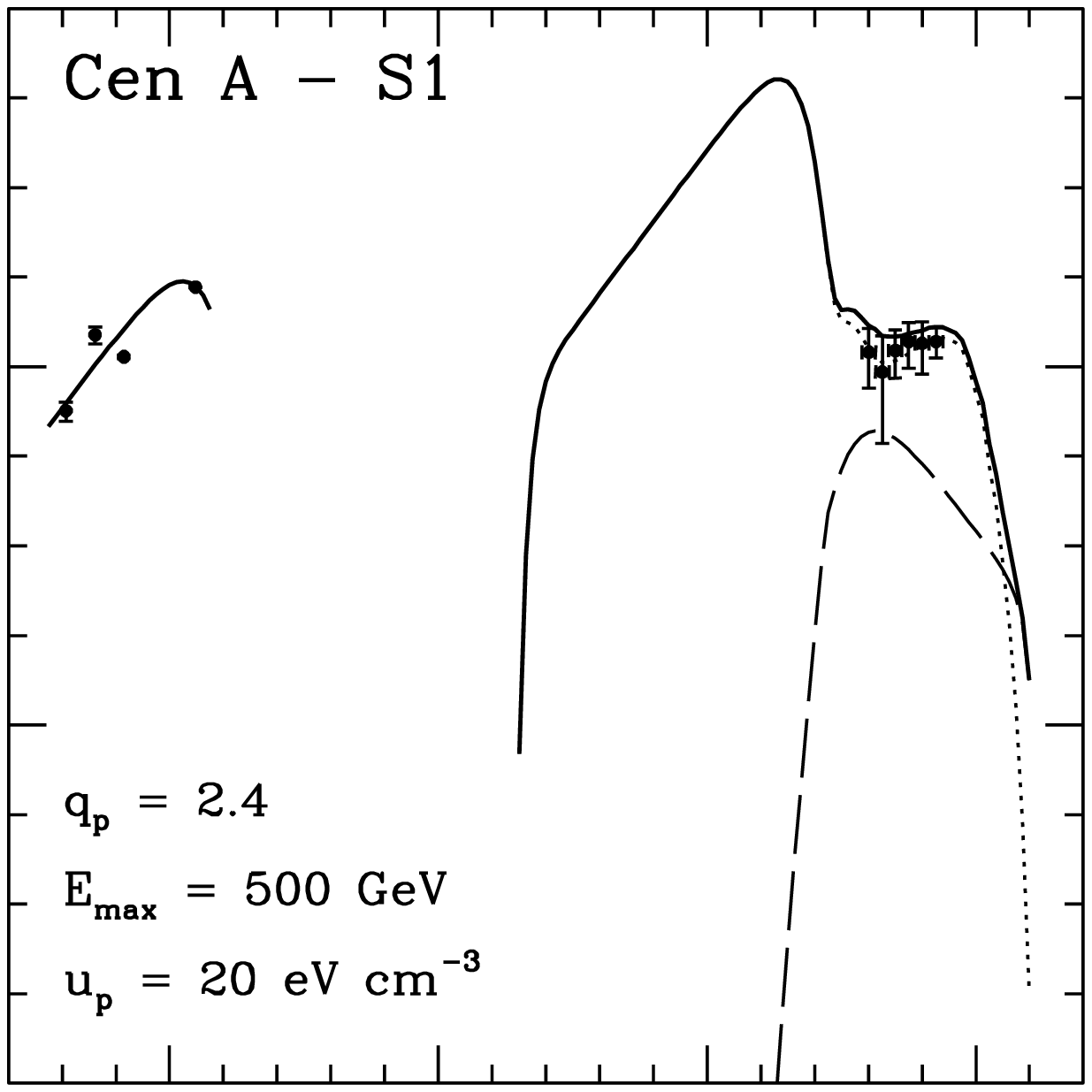}
\includegraphics{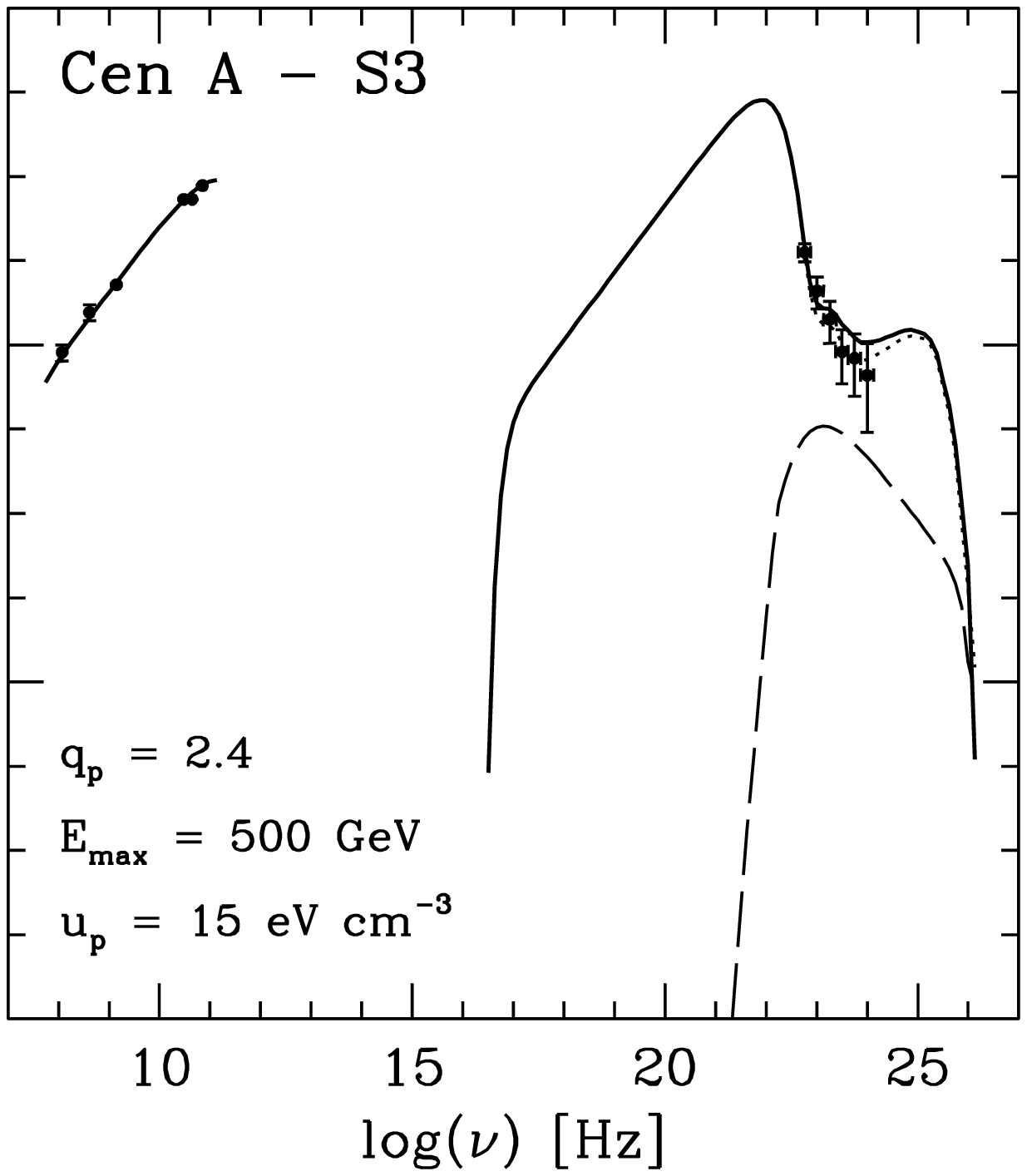}
\includegraphics{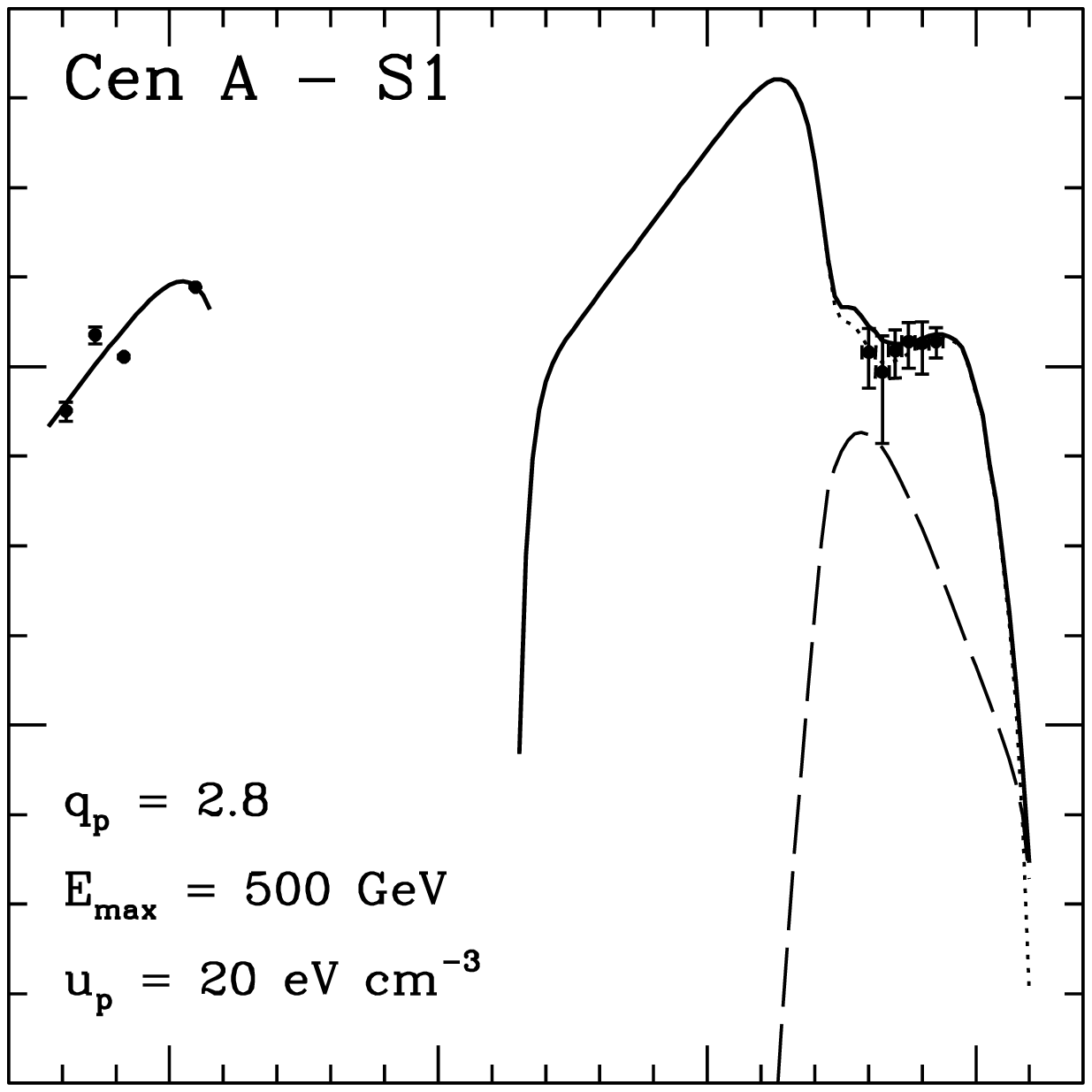}
\includegraphics{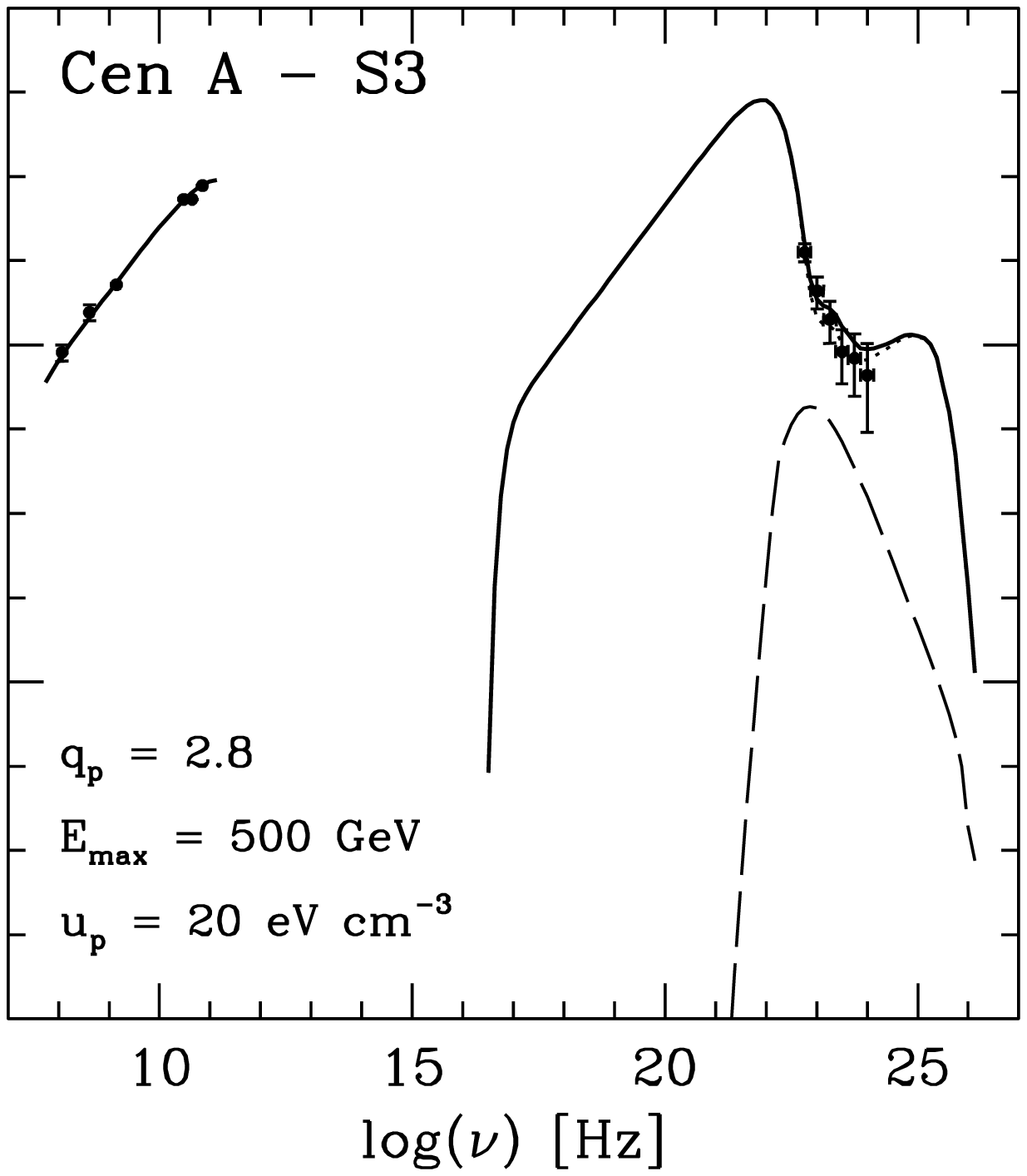}
\caption{
SEDs of two Cen\,A lobe regions S1, S3 with a lepto-hadronic model overalid on data. Emission components 
in the X-ray/$\gamma$-ray energy range are: superposed Compton, 
dotted line; pionic, dashed line; total: thick solid line. 
The leptonic components are as in Fig.\,\ref{fig:CenA_SED}. PED parameters are reported in each panel.
}
\label{fig:CenA_SED_leptohadronic}
\end{figure*}

The SEDs of the S1 and S3 regions of Cen\,A are sufficiently detailed so that the dominance of the Compton 
components provides a meaningful basis for spectral analysis to set bounds on the proton contents. We consider 
the two extreme cases, $q_p=2.0$ (flat hump) and $q_p=2.8$ (spiky hump) for each region (as shown in 
Fig.\,\ref{fig:CenA_SED_leptohadronic}). The {\it Fermi}-LAT data for S1 are interpreted as largely due to 
Compton scattering off the EBL, as is clearly 
reflected in the double-humped EBL profile. In a lepto-hadronic framework a substantial pionic hump tends to 
smooth out the Compton/EBL double hump feature if $q_p = 2.0$, and to enhance it unacceptably at low {\it 
Fermi}-LAT energies if $q_p=2.8$: in both cases, $u_p \mincir 20$ eV cm$^{-3}$. 
In a lepto-hadronic model for S3 a pion component contributes at the highest-energy ($q_p=2.0$) and lowest-energy 
($q_p=2.8$) {\it Fermi}-LAT data points: 
we infer, respectively, $u_p \mincir 10$ eV cm$^{-3}$ and $u_p \mincir 20$ eV cm$^{-3}$. 
In all cases lepto-hadronic models are inconsistent with the measurements for relatively high PED normalizations. 

\begin{figure}
\vspace{12.5cm}
\includegraphics{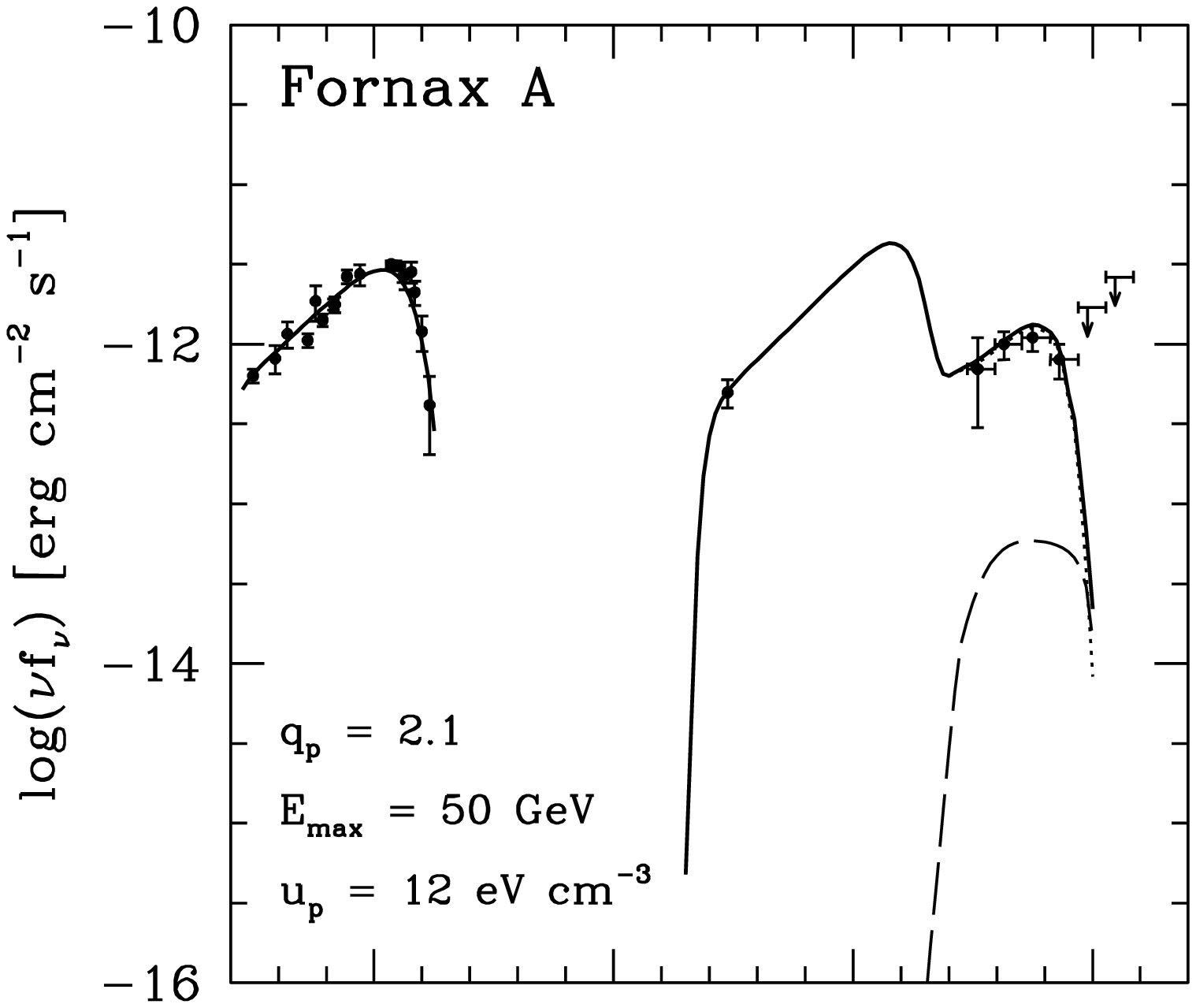}
\includegraphics{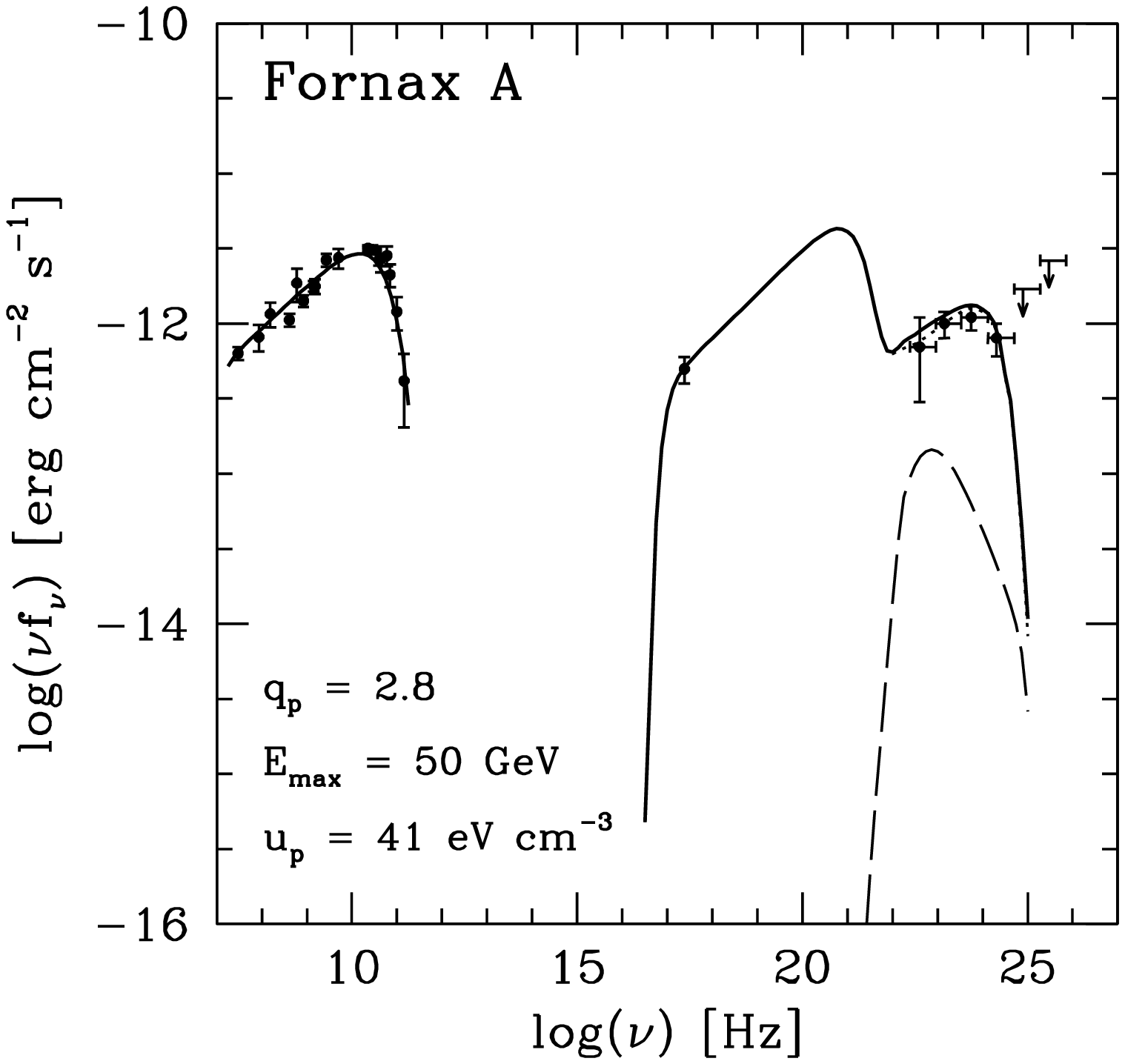}
\caption{ 
SEDs of Fornax\,A with a lepto-hadronic model overlaid on data. The leptonic component is as in Fig.\,2. 
PED parameters are indicated. Symbols are as in Fig.\,\ref{fig:CenA_SED_leptohadronic}. For the $q_p=2.1$ 
case see also Section 5.
}
\label{fig:FornaxA_SED_leptohadronic}
\end{figure}

By comparison,
the {\it Fermi}-LAT spectrum of Fornax\,A has a humpy shape similar to pionic emission from a PED with $q_p=2.1$ 
and $E_{\rm max} = 50$ GeV. The spectrum peaks at the third point, which has the smallest error bars; the above 
specified pionic emission peaks at similar energies. This coincidence leads to $u_p \mincir 12$ eV cm$^{-3}$, or 
possibly $u_p \mincir 40 $ eV cm$^{-3}$ (see Section 5). Steeper PED slopes make the pionic hump spikier and shift 
the pionic peak to lower energies; the larger uncertainty on the first {\it Fermi}-LAT point implies a weaker limit, 
$u_p \mincir 41$ eV cm$^{-3}$ for $q_p = 2.8$ (see Fig.\,\ref{fig:FornaxA_SED_leptohadronic}). 

For a given SED, limits are lower if $n_{\rm gas}$ is higher than the assumed nominal value. The number density of 
the thermal X-ray emitting gas in southern lobe of Cen\,A is $n_{\rm gas} = (0.9-2.5) \cdot 10^{-4}$ cm$^{-3}$ 
(Stawarz et al. 2013); for $n_{\rm gas} = 2.5 \cdot 10^{-4}$ cm$^{-3}$ the $u_p$ values would be a factor 2.5 lower 
than deduced above, e.g. $u_p < 4$ eV cm$^{-3}$ in S3 ($q_p=2$).

%
%

%

\section{Discussion} 

The electron energy density arises from $\gamma> \gamma_{min}$ SED-emitting electrons (index: $q_e$) and from $\gamma < 
\gamma_{min}$ electrons affected by non-radiative Coulomb losses (index: $q_e-1$; e.g. Rephaeli \& Persic 2015). Table\,1 
reports the electron and magnetic energy density values for the current lobes and for Fornax\,A (PR19). The electron and 
magnetic contributions are comparable in Fornax\,A and Cen\,A; in the latter system $u_e/u_B$ increases with increasing 
galactocentric radius. On the other hand, the magnetic contributions are relatively small in Cen\,B and NGC\,6251, as a 
consequence of the Compton humps in their SEDs being much more powerful than the synchrotron humps. 

Although radiatively inconspicuous, energetic protons in the lobes may dominate the NT energy density. In Section 3 we 
deduced upper limits on the proton content based on current {\it Fermi}-LAT data, $u_p \mincir 10$-40 eV cm$^{-3}$ 
depending on the lobe (region) and PED spectral index. These limits were obtained without accounting for uncertainties 
in the EBL and GFL photon densities. Uncertainties in the EBL density arise also from the fact that it is based on large 
datasets with inhomogeneous error estimates (in addition to other systematic uncertainties, such as redshift-dependent 
galaxy luminosity functions; Franceschini \& Rodighiero 2017). The main EBL components, the CIB and the COB, have $\sim$30\% 
uncertainties (Cooray 2016). 
A higher/lower COB power implies a lower/higher $N_{e,0}$ if the latter is determined via Compton/EBL modelling of 
{\it Fermi}-LAT data (e.g., the S1 region of Cen\,A); this in turn implies a higher/lower $B$ in the synchrotron fit to 
the radio spectrum
\footnote{
For example if $T_5$ is taken 9\% lower (to mimick a slightly lower COB), 
then in the Cen\,A S1 region $N_{e,0}$ and $B$ become, respectively, 
25\% higher and 10\% lower.
}
. The GFL suffers from observational uncertainties in the galaxy light distribution (Cen\,A: Dufour et al. 1979; Fornax\,A: 
Iodice et al. 2017), and from modeling uncertainties in the monochromatic to bolometric flux correction (e.g., Buzzoni et 
al. 2006) required to compute the dilution factor of the GFL blackbody model. A case in point is Fornax\,A, for which our 
estimate of the GFL photon field density (PR19) was based on nominal magnitude, color, and band-to-band magnitude conversion 
values adopted from the literature. The resulting nominal Compton/(EBL+GFL) flux slightly exceeds (by $\sim$10\%, yet within 
error bars) the nominal {\it Fermi}-LAT fluxes. As a result, the deduced UL is biased low; correcting for this leads to a weaker 
bound, $u_p \mincir 40$ eV cm$^{-3}$ for $q_p=2.1$. 

A lower limit to the proton content may be estimated assuming the lobe internal pressure to exceed the external (ambient) 
pressure. (Lobes are inflating bubbles in intergalactic space.) Since magnetic fields and relativistic electrons fall short 
of providing the required internal pressure, protons (NT and thermal) may provide the required pressure. For Cen\,A (see 
Table 1 for $u_e$ and $u_B$) this hypothesis leads to $u_p \magcir 5$ eV cm$^{-3}$ (Wykes et al. 2013). The thermal pressure 
inside the lobes is much smaller: in the S lobe $n_{\rm gas} = (0.9-2.5) \cdot 10^{-4}$ cm$^{-3}$ and $k_BT_{\rm 
gas} = 0.5$\,keV (Stawarz et al. 2013) imply a thermal energy density, $u_{\rm th} = \frac{3}{2} n_{\rm gas} k_BT$, in the 
range 0.068-0.188 eV cm$^{-3}$. So the above lower limit mainly applies to NT protons and is compatible with our ULs (see 
Fig.\,\ref{fig:CenA_SED_leptohadronic}).
\footnote{
No information on thermal gas is available for the lobes of Cen\,B and NGC\,6251. As to the latter, the lack 
of internal depolarization in the NW lobe implies, for a field strength of $B=0.4\,\mu$G and no field reversals 
in the lobe, $n_{\rm gas} < 0.75 \cdot 10^{-4}$ cm$^{-3}$ (see Willis et al. 1978). For Fornax\,A, $n_{\rm gas} 
= 3 \cdot 10^{-4}$ cm$^{-3}$ and $k_BT_{\rm gas} = 1$\,keV (W lobe; Seta et al. 2013) hence $u_{\rm th} = 0.45$ 
eV cm$^{-3}$. 
}

In spite of these uncertainties we may compare the estimated limits on $u_p$ with nominal values deduced from the proton 
to electron energy density ratio, $\kappa$, in an electrically neutral NT plasma where both particle species have energy 
distributions as specified in Section 3. For $q_p=2.0$, $q_e=2.2$ in Cen\,A's S1, S3 regions, $\kappa \simeq 120$, whereas 
for $q_p=2.1$, $q_e=2.43$ in Fornax\,A, $\kappa \simeq 110$ (Persic \& Rephaeli 2014). The corresponding values of $u_e$ 
(Table\,1) imply nominal values of $u_p$ that are compatible with our estimated ULs. 

The result $u_B + u_e + u_p > u_{\rm th}$ is likely to be a more general feature in the energetics of radio lobes. After 
all, lobes originate in the highly non-equilibrium phenomenon of AGN jets, and their evolution is very different from that 
of older galactic systems that have attained a state of hydrostatic or virial equilibrium.

\section{Conclusion} 

The SED analyses of Cen\,A, Cen\,B, and NGC\,6251 are quite similar to those of SYMA16, K13 and T12, respectively, 
who reported and modeled original data for these sources; for Cen\,A we adopted the separation of each lobe to three 
rectangular spatial regions defined by SYMA16. However, our treatment is appreciably different from those in 
previous works in the following respects: 
{\it (i)} we use a single EED, i.e. a truncated PL, throughout -- whereas SYMA16 used a generalized exponentially-cutoff 
PL that did produce analytically different EEDs in different lobe regions, whereas K13 and T12 used broken PLs. Doing so 
allowed us to make a more direct mutual comparison of the radiative properties of the different lobe regions; 
{\it (ii)} we use a more recent EBL model (Franceschini \& Rodighiero 2017, 2018; Acciari et al. 2019); and 
{\it (iii)} we fully account for the contribution of the host galaxy to the enhanced radiation field in the lobes. 
Our results confirm the main conclusions of the earlier analyses and favour a leptonic origin of the NT emission in the 
three lobe systems; for Cen\,A, whose SED is spatially resolved, a spectral evolution from the inner to the outer regions 
is clearly seen. The SEDs of the S1 and S3 regions of Cen\,A are sufficiently detailed for a spectral analysis to 
constrain the proton content to within few tens of eV cm$^{-3}$.

\section*{Acknowledgements}
This research has used the NASA/IPAC Extragalactic Database (NED), which is operated by the Jet Propulsion Laboratory, 
Caltech, under contract with NASA.


\begin{table*}
\caption[] {Energy densities (eV\,cm$^{-3}$) in the lobes.}

\begin{tabular}{ c  c  c  c  c  c  c  c  c  c  c}
\hline
\hline

\noalign{\smallskip}
Energy density  & & Cen\,A-N1 & Cen\,A-N2 & Cen\,A-N3 & Cen\,A-S3 & Cen\,A-S2 & Cen\,A-S1 & Cen\,B & NGC\,6251 & Fornax\,A \\
\noalign{\smallskip}
\hline
\noalign{\smallskip}
$u_e$           & &   0.043   &   0.041   &   0.028   &   0.026   &   0.058   &   0.061   &  4.623 &   0.121   &   0.404   \\ 
$u_B$           & &   0.022   &   0.036   &   0.060   &   0.099   &   0.039   &   0.022   &  0.021 &   0.004   &   0.209   \\ 
$u_e / u_B$     & &    2.02   &    1.15   &    0.48   &    0.26   &    1.49   &    2.79   &  224.5 &    30.5   &    1.93   \\ 
\noalign{\smallskip}

\hline
\hline

\end{tabular}

\end{table*}



\begin{thebibliography}{\small}

\bibitem{b0}  Abdallah H. et al. (H.E.S.S. Collab.), 2018, A\&A, 619, A71 
\bibitem{b1}  Abdo A.A. et al. ({\it Fermi}-LAT Collab.), 2010, Science, 328, 725
\bibitem{b2}  Acciari V., et al. (MAGIC Collab.), 2019, MNRAS, 486, 4233
\bibitem{b3}  Ackermann M. et al. ({\it Fermi}-LAT Collab.), 2016, ApJ, 826:1
\bibitem{b4}  Blandford R., Meier D., Readhead A., 2018, arXiv, 1812.06025
\bibitem{b5}  Bregman J.N., Snider B.A., Grego L., Cox C.V., 1998, ApJ, 499, 670
\bibitem{b6}  Buzzoni A., Arnaboldi M., Corradi R.L.M., 2006, MNRAS, 368, 877
\bibitem{b7}  Cooray A., 2016, Royal Society Open Science, 3, 150555
\bibitem{b8}  Crane P., Stiavelli M., King I.R., et al., 1993, AJ, 106, 1371
\bibitem{b9}  Day G.A., Thomas B.M., Goss W.M., 1969, Aust. J. Phys. Astrophys. Suppl., 11, 11
\bibitem{b10}  Dermer C.D., Menon G, 2009, High Energy Radiation from Black Holes (Princeton, NJ: Princeton University Press)
\bibitem{b11}  Dickey J.M., Lockman F.J., 1990, ARA\&A, 28, 215
\bibitem{b12} Dufour R.J., van den Bergh S., Harvel C.A., et al., 1979, AJ, 84, 284
\bibitem{b13} Finlay F.A., Jones B.B., 1973, Aust. J. Phys., 26, 389
\bibitem{b14} Franceschini A., Rodighiero G., 2017, A\&A, 603, A\,34 
\bibitem{b15} Franceschini A., Rodighiero G., 2018, A\&A, 614, C\,1 
\bibitem{b16} Franceschini A., Rodighiero G., Vaccari M., 2008, A\&A, 487, 837
\bibitem{b17} Gil de Paz A., Boissier S., Madore B., et al., 2007, ApJS, 173, 185
\bibitem{b18} Golombek D., Miley G.K., Neugebauer G., 1988, AJ, 95, 26
\bibitem{b19} Goss W.M., Shaver P.A., 1970, Aust. J. Phys. Astrophys. Suppl. 14, 1
\bibitem{b20} Hardcastle M.J., Cheung C.C., Feain I.J., Stawarz \L, 2009, MNRAS, 393, 1041
\bibitem{b21} Harris G.L.H., Rejkuba M., Harris W.E., 2010, PASA, 27, 457
\bibitem{b22} Haslam C.G.T., Salter C.J., Stoffel H., Wilson W.E., 1982, A\&AS, 47, 1 
\bibitem{b23} Israel F.P., 1998, A\&A Rev, 8, 237
\bibitem{b24} Jones P.A., Lloyd B.D., McAdam W.B., 2001, MNRAS, 325, 817
\bibitem{b25} Katsuta J., Tanaka Y.T., Stawarz \L., et al., 2013, A\&A, 550, A66 (K13)
\bibitem{b26} Kesteven M.J.L., 1968, Aust. J. Phys., 21, 369
\bibitem{b27} Komessaroff M.M., 1966, Aust. J. Phys., 19, 75
\bibitem{b28} Kraft R.P., Vazquez S.E., Forman W.R., et al., 2003, ApJ, 592, 129
\bibitem{b29} Krawczynski H., B\"ottcher M., Reimer A., 2012, in Relativistic Jets from Active Galactic Nuclei 
              (eds. M. B\"ottcher, D.E. Harris, H. Krawczynski; Berlin: Wiley), p.215
\bibitem{b30} Mack K.-H., Klein U., O'Dea C.P., Willis A.G., 1997, A\&AS, 123, 423
\bibitem{b31} Mathewson D.S., Healey J.R., Rome J.M., 1962, Aust. J. Phys., 15, 354
\bibitem{b32} McAdam W.B., 1991, Proc. Astron. Soc. Aust., 9, 255
\bibitem{b33} McKinley B., Briggs F., Gaensler B.M., et al., 2013, MNRAS, 436, 1286
\bibitem{b34} McKinley B., Yang R., L\'opez-Caniego M., et al., 2015, MNRAS, 446, 3478
\bibitem{b35} Mills B.Y., 1952, Aust. J. Sci. Res. 5, 266
\bibitem{b36} Mills B.Y., Slee O.B., Hill E.R., 1961, Aust. J. Phys., 14, 497
\bibitem{b37} Milne D.K., Hill E.R., 1969, Aust. J. Phys., 22, 211
\bibitem{b38} O'Sullivan S.P., Feain I.J., McClure-Griffiths N.M., et al., 2013, ApJ, 764:162
\bibitem{b39} Persic M., Rephaeli Y., 2007, A\&A, 463, 481 
\bibitem{b40} Persic M., Rephaeli Y., 2014, A\&A, 567, A\,101
\bibitem{b41} Persic M., Rephaeli Y., 2019, MNRAS, 485, 2001 (PR19)
\bibitem{b42} Rephaeli Y., Persic M, 2015, ApJ, 805:111
\bibitem{b43} Romanishin W., 1986, AJ, 91, 76
\bibitem{b44} Sanders D.B., Mirabel I.F. 1996, ARAA, 34, 749
\bibitem{b45} Schreier E.J., Feigelson E., Delvaille J., et al., 1979, ApJ 234, L39
\bibitem{b46} Schr\"oder A.C., Mamon G.A., Kraan-Korteweg R.C., Woudt P.A., 2007, A\&A, 466, 481
\bibitem{b47} Seta H., Tashiro M.S., Inoue S., 2013, PASJ, 65, 106
\bibitem{b48} Shaver P.A., Goss W.M., 1970, Aust. J. Phys. Astrophys. Suppl., 14, 133
\bibitem{b49} Stawarz \L., Tanaka Y.T., Madejski G., et al., 2013, ApJ, 766:48
\bibitem{b50} Sun X.-N., Yang R.-Z., McKinley B., Aharonian F., 2016, A\&A, 595, A\,29 (SYMA16)
\bibitem{b51} Takeuchi Y., Kataoka J., Stawarz \L., et al., 2012, ApJ, 749:66 (T12)
\bibitem{b52} Tashiro M., Kaneda H., Makishima K., et al., 1998, ApJ, 499, 713
\bibitem{b53} Willis A.G., Wilson A.S., Strom R.G. 1978, A\&A, 66, L1
\bibitem{b54} Yang R.-Z., Sahakyan N., de Ona Wilhelmi E., Aharonian F., Rieger F., 2012, A\&A, 542, A\,19
\bibitem{b55} Wilson A.S., Smith D.A., Young A.J., 2006, ApJ, 644, L9
\bibitem{b56} Wykes S., Croston J.H., Hardcastle M.J., et al., 2013, A\&A, 558, A19
\end{thebibliography}
\end{document}